\documentclass{easees2013}
\usepackage{graphicx}
\usepackage[authoryear]{natbib}
\usepackage[fleqn]{amsmath}
\usepackage[]{amsfonts}%
\usepackage{xcolor}
\usepackage{amssymb}
\usepackage{url}

\newcommand{\apj}{ApJ}
\newcommand{\apjs}{ApJS}
\newcommand{\apjl}{ApJ Letters}
\newcommand{\aap}{A\&A}
\newcommand{\aaps}{A\&AS}
\newcommand{\mnras}{MNRAS}
\newcommand{\nat}{Nature}

\newcommand{\araa}{ARA\&A}
\newcommand{\sovast}{Sov. Ast.}
\newcommand{\memsai}{Mem. Soc. Ast. Ital.}
 
\newcommand{\apss}{Ap\&SS}

\newcommand{\pasj}{Pub. Ast. Soc. Japan}

\newcommand{\revmodphys}{Rev. Mod. Physics}
\newcommand{\solphys}{Sol. Phys.}

\DeclareGraphicsExtensions{.pdf}

\newcommand{\Gyr}{Ga}

\TitreGlobal{The Ages of Stars}
\begin{document}

\title{How accurate are stellar ages\\[0.3cm]
 based on stellar models?\\[0.6cm]
II. The impact of asteroseismology}  

\runningtitle{Y. Lebreton, M.J. Goupil \& J. Montalb\'an:
Ages and Asteroseismology}%

\author{Y. Lebreton}\address{Observatoire de Paris, GEPI, 
CNRS UMR 8111, 5 Place Jules Janssen, 
92195 Meudon, 
France, and Institut de Physique de Rennes, Universit\'e de Rennes 1, 
CNRS UMR 6251, 35042 Rennes, France. Email: \url{yveline.lebreton@obspm.fr}}
\author{M.J. Goupil}\address{Observatoire de Paris, LESIA, CNRS UMR 8109, 
92195 Meudon, France.\\ Email: \url{MarieJo.Goupil@obspm.fr}}
\author{J. Montalb{\'a}n}\address{Email: \url{j.montalban@skynet.be}}%

\begin{abstract}
Accurate and precise stellar ages are best determined for stars which are
strongly observationally constrained, that is which are intrinsically oscillating. We review here 
the seismic diagnostics which are sensitive to  stellar ages and provide some illustrating examples of
seismically age-dated stars.
\end{abstract}
\maketitle

\section{Introduction}

\begin{figure}
\begin{center}
\includegraphics[width=0.8\textwidth]{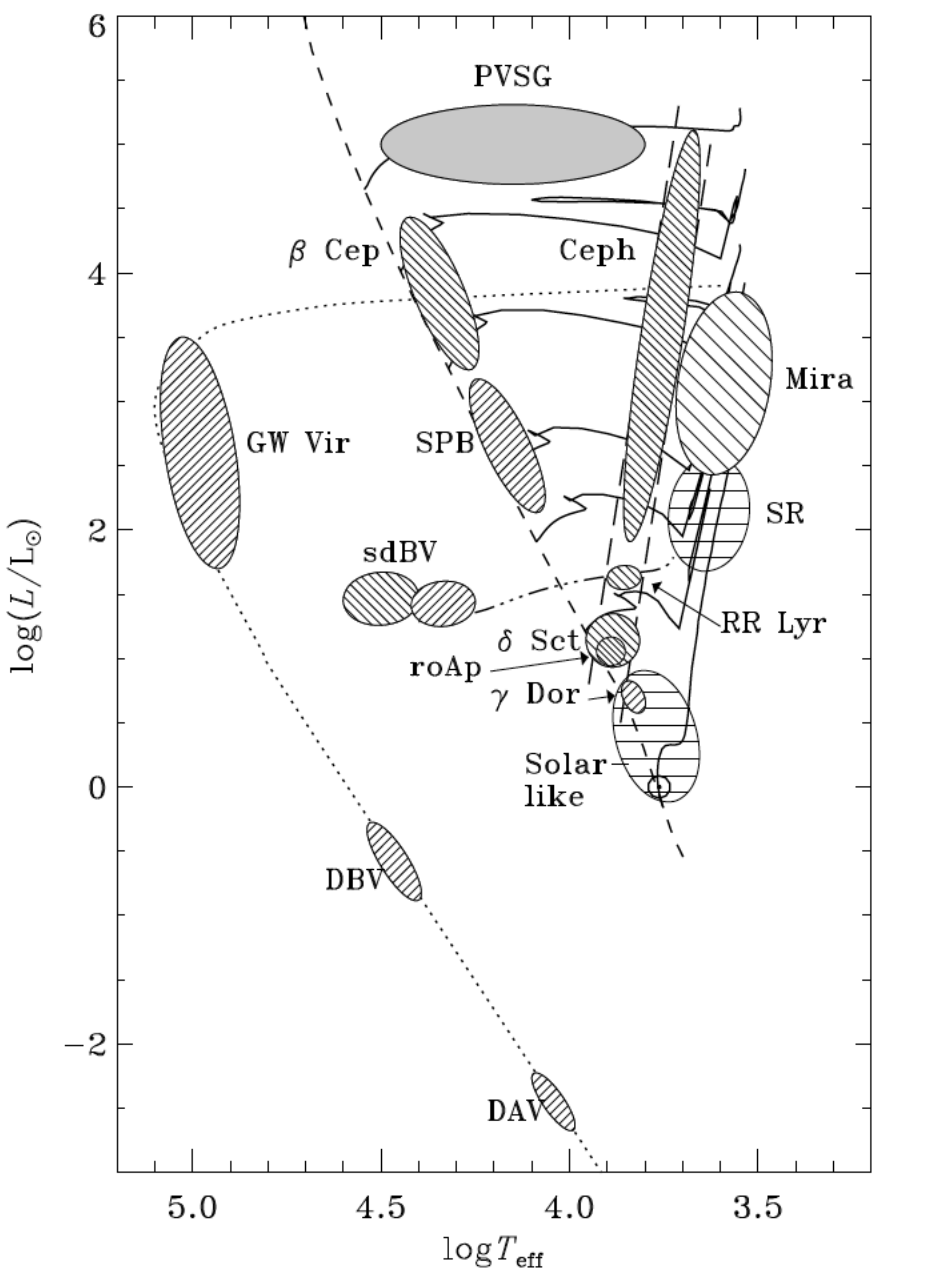}
\caption{Regions in the HR diagram where oscillating
 stars are observed. 
 [From J. Christensen-Dalsgaard,
{\protect\url{http://astro.phys.au.dk/~jcd/oscilnotes/contents.html}.]}
}
\label{HRpuls}
\end{center}
\end{figure}

Precise and absolute stellar ages are crucial requirements in many astrophysical studies. 
In particular, the huge harvest of newly discovered exoplanets calls for accurate and precise  ages of their host-stars, a crucial 
parameter in the 
understanding of planet formation and evolution \citep[][see also the 
chapter by T. Guillot]{2011A&A...531A...3H}.

As we discussed in Part I, the determination of the age of a single star is very imprecise when the only observational data 
available for that star
are its luminosity, effective temperature, and metallicity. Fortunately, there are several kinds of stars for which additional 
observational data are 
available, allowing to better constrain stellar models and thus to improve age-dating. Stellar masses can be obtained directly, via 
the third Kepler 
law, for stars in binary systems (visual/interferometric binaries or combined spectroscopic/eclipsing binaries). On the other hand, 
stellar radii are 
accessible for stars in combined eclipsing/spectroscopic binaries or, for bright or large stars observed in interferometry. 
Furthermore, major steps 
towards the precise and accurate age-dating (and weighing) of oscillating stars have been made after the asteroseismic data 
provided by the ultra high precision photometric space-borne missions 
 \textsl{CoRoT} \citep{2002esasp.485...17b} and \textsl{Kepler} 
\citep{2010ApJ...713L..79K} have been made available.

As illustrated in Fig.~\ref{HRpuls}, stellar pulsations are observed in stars spanning the whole Hertzsprung-Russell (HR) diagram. 
The  analysis of the observed frequencies allows to probe the physical processes that are at work 
in the propagation regions at the actual age of  the star, but also that were at work in its past life. This was demonstrated a very long time ago, by the study of large amplitude, radial, classical pulsators as 
Cepheids \citep[starting with the work of ][]{1917Obs....40..290E}  and by helioseismology, following the discovery of low-amplitude pulsations in the 
Sun by \citet{1962ApJ...136..493E} and \citet{leigh62}.
The field of asteroseismology slowly evolved after the first observation of
 low amplitude solar-like oscillations in Procyon \citep{1991ApJ...368..599B},  but  made significant advances 
 with the advent of ultra-high precision photometry conducted in space by \textsl{CoRoT} and \textsl{Kepler}.
Indeed, these missions have detected low amplitude stellar oscillations in many stars allowing to measure their frequencies with a typical 
precision of a few tenths of micro Hertz \citep[\eg,][]{2008Sci...322..558M,2013ARA&A..51..353C}.

Seismic data have been used to age-date and weigh many stars. Currently, two approaches are taken. The first one, ensemble 
asteroseismology, attempts to determine the age and mass of large sets of stars based on their mean seismic properties \citep{2014ApJS..210....1C}. 
In this approach, interpolation in large grids of ``ready-made'' stellar models, by different techniques, provides the mass and age of the model that 
best matches the observations. The alternative approach, more precise, is the hereafter named ``\`{a} la carte'' modelling \citep{2013eas63123}, 
that is the detailed study of  specific stars, one by one, also referred to as ``boutique''  modelling \citep[see \eg,][]{soderblom13}.
 This approach has been used to model \textsl{CoRoT} and \textsl{Kepler} stars \citep[see the reviews by][]{2013ASPC..479..461B,2013ARA&A..51..353C}.
In particular, stars hosting exoplanets have been modelled by for instance \citet{2011ApJ...726....2G}, \citet{2012ASPC..462..469L}, 
\citet{ 2012A&A...543A..96E}, \citet{2013eas63123}, \citet{2013ApJ...766...40G}, and \citet{2014arXiv1406.0652L}.

In this lecture, we focus on asteroseismic diagnostics of stellar ages and also briefly discuss the impact of
 asteroseismology in stellar mass and radius determination. We mainly restrict the subject to low-amplitudes oscillations in solar-like oscillators, 
that is damped oscillations stochastically excited by turbulence in the near surface layers. We 
 concentrate on the advantages brought by the calculation of \`{a} la carte  stellar models with respect to ensemble asteroseismology. 
The idea is to perform \`{a} la carte modelling of the stars with the strongest and more numerous observational constraints, 
which we refer to as calibrators.  Calibrator modelling is expected to provide new insights and constraints on the complex physical 
processes at work in stellar interiors as well as on some of the unknown or -at least- very imprecise properties of any star, 
as its initial helium content. As discussed in details in Lecture 1, the poor knowledge we have of these aspects is responsible for 
large uncertainties in the age-dating of individual stars. The characterization of calibrators will then serve to improve the modelling and age-dating of other, less well-observed stars, in particular for statistical studies (for instance for Galactic structure and evolution studies).  In the future, one can expect that a sufficient 
number of calibrators is known, paving the HR diagram in all its dimensions (mass, age/evolutionary stage, and chemical composition). 
Therefore, to any ordinary star will correspond a referent calibrator to be used to improve its characterization.

This chapter is organized as follows. In Section \ref{alacarte}, we give a 
definition for \`{a} la carte stellar models and present some examples. In
 Section~\ref{oscil}, we briefly present oscillation frequency calculations 
and we detail the different seismic diagnostics that are currently used to 
constrain stellar models and their ages. In Section~\ref{applications}, we
 present selected examples of age-dating based on asteroseismology and 
discuss the range of precision and accuracy that can be reached.

\section{``\`{A} la carte'' stellar modelling}
\label{alacarte}

In \`{a} la carte modelling, dedicated internal structure models are built to best represent the observed star.
Given the observational data available for that star, an optimization process is conducted to define and calculate 
the stellar model that best matches these data. In the following, we distinguish so-called {\sl classical} 
observational parameters from {\sl seismic} ones. Classical parameters are, for instance, the effective temperature 
$T_\mathrm{eff}$, the luminosity $L$, the metallicity [Fe/H] or the individual abundances, the surface gravity $\log g$, 
and when measurable, the mass $M$, the radius $R$, the mean density derived from the transit of an exoplanet, etc. 
Seismic parameters can be the individual oscillation frequencies or different combinations of them (see Sect.~\ref{diagnos}). 
In the optimization process, an adjustment of the input physics and other parameters of the stellar models (\eg, the initial helium abundance) 
is performed, to match at best the observations. Naturally, the more observational constraints, the more model parameters can be adjusted, as 
illustrated in the following.

The modelling of calibrators (Sun, very well-observed stars) is mainly performed \`{a} la carte. 
As recalled below, the solar model and other calibrator models provide the reference for modelling other stars, 
in terms of input physics, helium abundance, etc. Reference models also serve to calibrate empirical relations 
(for empirical age-dating, see the chapter by R. Jeffries). \`{A} la carte stellar models are also required when 
a precise characterization of an individual star is required, for instance when absolute ages are demanded. 
This is the case of exoplanet host-stars, for which we need the age, mass, and radius to characterize the exoplanet 
internal structure, formation scenario, and evolution (see the chapter by T. Guillot).

\subsection{The Sun}
\label{Sun}

The first calibrator ever studied is the Sun, see, for instance, the first paper on the helium content of the Sun by
 \citet{1946ApJ...104..203S} where he 
found a mass fraction of helium $Y\approx 0.40$ and suggested an upper limit on the solar 
metallicity $Z/X$ of 0.255, \ie, about ten times the present one, inferred from both the solar spectrum and analysis of 
meteorites! The solar model is a reference and provides tests and calibration for the physical description 
and free parameters of stellar models. In the case of the Sun, the classical parameters (mass $M_\odot$, radius $R_\odot$, 
luminosity $L_\odot$, individual abundances of chemical elements, age $t_\odot$) are very precisely measured. One commonly 
defines the {\sl standard solar model} as the model of one solar mass that matches, at solar age, the observed solar luminosity, 
radius, and surface metallicity. The standard solar model is based on standard input physics, that is currently accepted and updated 
physical inputs at the time of the calculation. Once the input physics has been fixed, the calibration provides an estimate for the 
two main free parameters of the model, the initial helium content, which mainly controls the luminosity, and the mixing-length 
parameter of convection $\alpha_\mathrm{conv}$, which mainly controls the 
radius \citep[see for instance works by][]{1979A&A....79..251M, 
1980A&A....84..135H,1982MNRAS.199..735C,1986A&A...161..119L}. A few decades ago, the so-called solar neutrino problem 
bothered solar modellers \citep[see \eg,][]{1968ApJ...153..113B} and some of them designed so-called non-standard solar models 
in order to reproduce the observed neutrino flux, see for instance the solar models with turbulent diffusion by 
\citet{1981A&A....96....1S, 1987A&A...175...99L}, or the exotic solar model with a central black hole by \citet{1973ApL....13...45S}. 
This was prior to the finding that neutrinos have a mass and oscillate in three flavours in their travelling from Sun to Earth. 

 \begin{figure}
 \begin{center}
\includegraphics[width=0.8\textwidth]{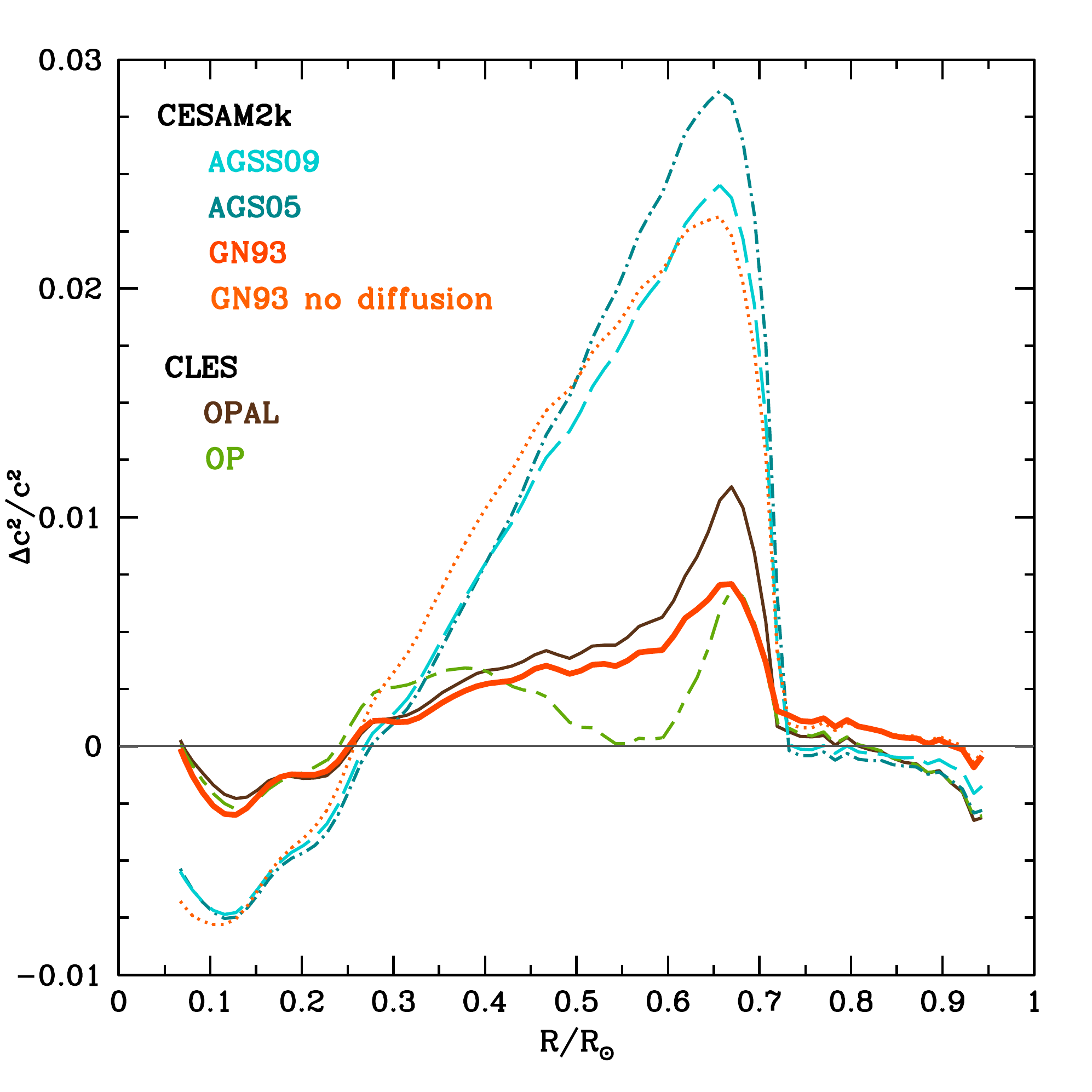}
 \caption{Relative differences between the sound speed derived through inversion
of oscillation frequencies by \citet{2000ApJ...529.1084B} and the
sound speed obtained in different solar models. The inputs of solar models
 are discussed in Part I.
[From Y. Lebreton, in the collective book by \citet{2012EAS....54....3S}.] 
 }
 \label{solarsoundspeed}
 \end{center}
 \end{figure}
 
Very importantly, a major step towards the resolution of the solar neutrino problem came for helioseismology. Indeed, the very
 complete and precise set of solar pressure mode oscillation frequencies (see Sect.~\ref{oscil}) provided by radial velocity and intensity measurements  brought strong 
  clues that the standard solar model, even imperfect, was not so far from reality \citep[see \eg, the review 
 by][]{1988RvMP...60..297B}. Heliosismic analysis allowed to determine the present helium abundance in the solar convective 
 envelope \citep[\eg,][]{1984MmSAI..55...13G, 1984LIACo..25..264D,1995MNRAS.276.1402B}  as well as the depth of this 
 envelope 
 \citep{1991SvA....35..507K,1991ApJ...378..413C}. In turn, this brought constraints on the input physics to be included in a solar 
 model, mainly the gravitational settling of chemicals \citep{1996A&A...312.1000R} and elaborated equations of state 
 \citep{1988Natur.336..634C}.
 However, as illustrated in Fig.~\ref{solarsoundspeed}, there remained disagreements between the sound speed profile inferred 
 from the inversion of the 
 observed frequencies and the one obtained in solar models,
 in particular in the radiative regions just below the convective envelope and in the centre. Maximum relative 
 differences in the square of the sound speed,
 $(c^2_\mathrm{models}-c^2_\mathrm{observed})/c^2_\mathrm{observed}$ 
 are of $\sim 0.4$ per cent below the convective zone
 \citep[\eg,][]{1996Sci...272.1286C}. These differences can partially be erased by the introduction of a transport of chemical elements 
 in the radiative zone \citep{1997A&A...327..771G}. 
 More recently, \citet{2012ApJ...746...50Z} showed that  the disagreement in the sound speed between
 solar models and helioseismic inversions below the base 
 of the solar convective envelope nearly disappears when diffusive overshooting mixing is taken into
 account  at the transition between the  upper convective layers and the lower radiative 
 interior \citep[see also][]{2011MNRAS.414.1158C}.
 
 Helioseismology still calls for further improvements and refinements in the solar model description, regarding for 
 instance internal rotation \citep[see][]{1998ApJ...505..390S,1999MNRAS.308..405C}, and the description of the tachocline \citep[see][for a review]{2011ApJ...733...12R}. 
 Furthermore, the redetermination of the solar abundances on the basis 
 of  3-D model atmospheres and improved atomic data led to a degradation of the quality of the 
 modelling  of the sound speed profile. With the new \textsl{AGSS09} abundances (presented in Lecture 1), the difference  
 $(c^2_\mathrm{models}-c^2_\mathrm{observed})/c^2_\mathrm{observed}$ 
 is about $10$ times larger than in the \citet{1996Sci...272.1286C} solar model based on the 
  \textsl{GN93} mixture \citep[see \eg,][and Fig.~\ref{solarsoundspeed}]{2009ara&a..47..481a}. 
Finally, gravity modes which would allow to probe the deep interior of the Sun
 (see Sect.~\ref{oscil} below)  are still
 tracked, but their low amplitudes make the quest difficult \citep[\eg,][]{2007Sci...316.1591G}.

\subsection{Other stars}

For other stars, we are quite far from the degree of achievement of the solar model. 
We actually assume solar values for the free parameters
entering their physical description,
even if they may have masses, chemical compositions, and evolutionary stages far from  solar. 
However, when the observational data are very precise and numerous,  the status of {\sl calibrator} 
can be given to a star. Then, with \`{a} la carte modelling, we hope to reach a level 
where we are able to discriminate different options for the input physics of stellar models and get 
insight on mostly free parameters, as the helium content. For stellar {calibrators}, the age determination 
will therefore be not only more precise, but also more accurate. We give some examples later in the text.

\subsection{Optimization procedures}

In seismic \`{a} la carte modelling, the constraints on the classical observed stellar parameters 
and those on the seismic ones are
 combined to improve the calibration of the model of the considered star. In the examples presented in Sect.~\ref{applications}, 
 $L$, $T_\mathrm{eff}$, [Fe/H] will be considered together with different seismic indicators. 
 Then, stellar models are calculated and their
unknowns, which may be either inputs or outputs of the model calculation, are adjusted so as to minimize the deviation of the model from
the observational material. 
 The goodness of the match can be evaluated through the minimization of a merit function $\chi^2$, which can be written,
\begin{equation}
\label{chi2}
\chi^2 \; = \; 
\sum_{i=1}^{N_\mathrm{obs}} \frac{\left ( x_\mathrm{i, mod} -
 x_\mathrm{i, obs} \right ) ^2}{\sigma_\mathrm{i, obs}^2} \, , 
\end{equation}
where $N_\mathrm{obs}$ is the total number of observational constraints considered, 
$x_\mathrm{i, mod}$ and  $x_\mathrm{i, obs}$ are the computed and observed values of 
the $i^\mathrm{th}$ constraint.
The more observational constraints available, 
the more free parameters can be adjusted in the modelling process. The parameters that are currently adjusted are the initial 
stellar mass $M$, the initial $(Z/X)_0$ and helium mass fraction $Y_0$ (all are models inputs), the parameters entering the 
physical 
description of the star (mainly the mixing-length parameter of convection $\alpha_\mathrm{conv}$ and overshooting parameter 
$\alpha_\mathrm{ov}$), and the age $A$ (model output). If too few observational constraints are available, 
some free parameters have to be fixed (see below), while the others can be adjusted. Note that the expression of the $\chi^2$ in 
Eq.~\ref{chi2} is not valid if data are correlated. In that case, Eq.~\ref{chi2cor} has to be used (see Sect.~\ref{correl}).

Different optimizations approaches have been used to adjust stellar models to observational constraints: 
 model-grid based methods \citep[][etc.]{2009ApJ...700.1589S, 2010ApJ...725.2176Q}, genetic algorithms as in the 
Asteroseismic Modelling Portal \citep[][]{2009ApJ...699..373M}, 
Levenberg-Marquardt minimization 
\citep{2005a&a...441..615m}, and Bayesian approach \citep{2012ApJ...749..109G, 2012MNRAS.427.1847B}. The 
Levenberg-Marquardt (LM) algorithm combines the steepest descent method and Newton's method to solve non-linear least 
square problems \citep{2003drea.book.....B,2002nrc..book.....p}. In some examples presented below the LM minimization 
method was used in the way described by \citet[][]{2005a&a...441..615m}. 

\section{Stellar oscillations}
\label{oscil}
\subsection{Setting the stage}

Pulsating stars are spanned all across the HR diagram (Fig.~\ref{HRpuls}). 
There are many categories of oscillators both in terms of the kind of pulsations observed (radial or non-radial),  
and of their amplitudes and driven mechanism. In particular, one distinguishes pulsating stars where the amplitudes 
of the oscillations are large (Cepheids, Mira, etc.) from low-amplitude pulsators (solar-like, $\beta$ Cephei, etc.). 
The characteristic oscillation period $P$ scales as the dynamical time-scale $t_\mathrm{dyn}$, 
which measures the amount of time it would take a star to collapse in the absence of any internal pressure. Therefore $P$ reads,
\begin{equation}
P\approx t_\mathrm{dyn} \; = \; \left( \frac{2 R^3}{G M}\right)^{1/2}
 \propto \langle \rho \rangle^{-1/2} \quad .
\end{equation}

The theory of stellar oscillations is presented and detailed in text books, see \eg,
\citet[][]{1980tsp..book.....C}, \citet{1989nos..book.....U}, 
\citet{CoursJCD}\footnote{{\protect\url{http://astro.phys.au.dk/~jcd/oscilnotes/contents.html}}},
\citet{2010aste.book.....A}, and references therein. In the following, we only briefly recall the main steps of 
the calculation of stellar oscillations. We focus on  solar-like oscillations, excited by stochastic convective motions
 that take place in low-mass stars. The related oscillators are multi-mode oscillators.

\subsection{Stellar oscillations equations}

The structure of a star is described with the classical equations of hydrodynamics,
\begin{equation}
\frac{\partial \rho}{\partial t}+\overrightarrow{\nabla}.(\rho 
\vec{u}) \; = \; 0,\ \hfill\ \mathrm{continuity}
\end{equation}

\begin{equation}
\rho\left (\frac{\partial }{\partial t}+\vec{u}.
\overrightarrow{\nabla}  \right )\vec{u} \; = \; \rho\vec{f}-\overrightarrow{\nabla}$P$-\rho\overrightarrow{\nabla}\Phi+ \mathop{\mathrm{div}}\mathfrak{S},\ \hfill\ \mathrm{momentum}
\end{equation}

\noindent with
\begin{equation}
\Delta \Phi \;  = \; {\nabla}^2 \Phi = 4\pi G \rho, \hfill \mathrm{Poisson's\ equation}
\end{equation}

\begin{equation}
\rho T\left (\frac{\partial }{\partial t}+\vec{u}.
\overrightarrow{\nabla}  \right ){S} \; = \; \rho (\epsilon_\mathrm{nuc}+\epsilon_\mathrm{visc}) - \overrightarrow{\nabla}.\overrightarrow{F_\mathrm{R}},
 \hfill  \mathrm{energy\ conservation}
\end{equation}
where $\rho$ is the density, $P$ the pressure, $T$ the temperature, $\vec{u}$ the velocity of the flow, 
$\vec{f}$ the external forces, $\Phi$ the gravitational potential, $\mathfrak{S}$ the viscous stress tensor,  $\overrightarrow{F_\mathrm{R}}$ the heat flux, $S$ the entropy, and $\epsilon_\mathrm{nuc, visc}$ the energy produced or lost by nuclear reactions, neutrino loss, viscous heat generation, etc.

The basic assumptions consist in neglecting external forces, dissipation, and shear instabilities. 
Furthermore, 
oscillations occur on a dynamical
 time scale, which in the evolutionary phases of interest here is much 
 smaller than both the Kelvin-Helmholtz and nuclear time scales (see definitions in Lecture 1).   
Oscillations can  then be treated as small perturbations of the 
  equilibrium model at a given evolutionary stage.   Hereafter, the  variables defining 
  the equilibrium state of the star  are labelled with the subscript $0$ 
($\vec{u_0}=\vec{0}$, $p_0$, $\rho_0$, etc.). The equations are then linearised in the
   perturbations with a classical time dependency in $\exp(-i \omega t)$. 
    The linear perturbations associated with the oscillations 
are then defined as
\begin{equation}
\vec{\xi}(\vec{r},t)\; = \; \vec{\xi}(\vec{r})\, \exp(-i \omega t),
\end{equation}
where $\vec{\xi}(\vec{r})$ is the radial displacement vector. The
 velocity and pressure read,
\begin{equation}
\vec{u}(\vec{r},t) \; = \; -i \, \omega\ \vec{\xi}(\vec{r})\ 
\exp(-i \omega t)\ ; \ p(\vec{r}, t)=[p_{0} (\vec{r}) +
 p^{\prime}(\vec{r})] \exp(-i \omega t) \; ,
\end{equation}
where $p^{\prime}(\vec{r})$ is the pressure perturbation, etc.
The horizontal dependence of the functions is then expressed using the spherical harmonics, for instance  for the pressure perturbation,
\begin{equation}
p^{\prime}(\vec{r}) \; \propto \; p^{\prime}({r})\times Y_{\ell}^{m} (\theta, \phi),\ \mathrm{etc.}
\end{equation}
where $\theta$ and $\phi$ denote the usual spherical coordinates.
This provides an equation for the radial displacement 
\begin{equation}
-\omega^{2} \rho_{0} \xi_{r} + \dfrac{\mathrm{d} 
p^{\prime}}{\mathrm{d} r} + \rho_{0}\dfrac{\mathrm{d} 
\Phi^{\prime}}{\mathrm{d} r}+ \rho^{\prime} g_{0}=0 \; ,
\end{equation}
as well as associated equations for the perturbations of the Poisson's equation and continuity equation.

In a third step, the heat exchanges may be discarded, which does not impact the frequencies significantly. In such an adiabatic approximation 
\begin{equation}
\dfrac{p^\prime}{p_0} \; = \; \Gamma_{1} \dfrac{\rho^\prime}{\rho_{0}} \, ,
\end{equation}
 where 
 \begin{equation}
\Gamma_{1} \; = \; \left(\dfrac{\partial \ln p}{\partial \ln 
\rho}\right)_\mathrm{ad} \; ,
 \end{equation}
is the first adiabatic exponent.

The surface and central properties of a star allow standing waves to develop, which
are at the origin of the observed surface oscillations.  
 In the case of pressure modes for instance, modes are reflected  
at the surface because of the 
 decrease of the density, and are refracted on the inner boundary of the trapping 
cavity because of the inwards increase of the (adiabatic) sound speed 
 ${c}_\mathrm{s}=\sqrt{\Gamma_{1} p/\rho}$. 
Note that for an ideal gas, ${c}_\mathrm{s}\propto \sqrt{T/\mu}$, where $\mu$
is the mean molecular weight.
Mathematically,  boundary conditions must be imposed  when solving the oscillation equations; the problem of stellar 
oscillations is 
therefore an eigenvalue problem. The solution is a set of discrete frequencies $\nu_{n, \ell, m}=\omega_{n, \ell, m}/2\pi$, each 
frequency being characterized by three numbers (integers). 

The radial dependence of an eigenfunction is characterized by the number $n$, the radial order, which counts the number of 
nodes along a stellar radius.
The horizontal dependence is represented by the spherical harmonic $Y_{\ell, m}$. The number $\ell$ is the angular degree, it 
counts the number of nodal lines on the surface, while $m$ is the azimuthal order, which counts the number of nodal lines crossing the equator. In absence of rotation, modes of same $n$, $\ell$, and different $m$ have the same frequency. 
We neglect rotation in the following and therefore write $\nu=\nu_{n, \ell}$.

We point out that the observations of \textsl{CoRoT} and \textsl{Kepler} have provided sets of oscillation frequencies $\nu_{n, \ell}$ 
for many stars with a very high precision. These frequencies are distance-independent and provide very strong constraints for the modelling of oscillating stars.

\subsection{Nature of the oscillation modes}
\label{pgmode}
\begin{figure}
\begin{center}
\includegraphics[width=0.8\textwidth]{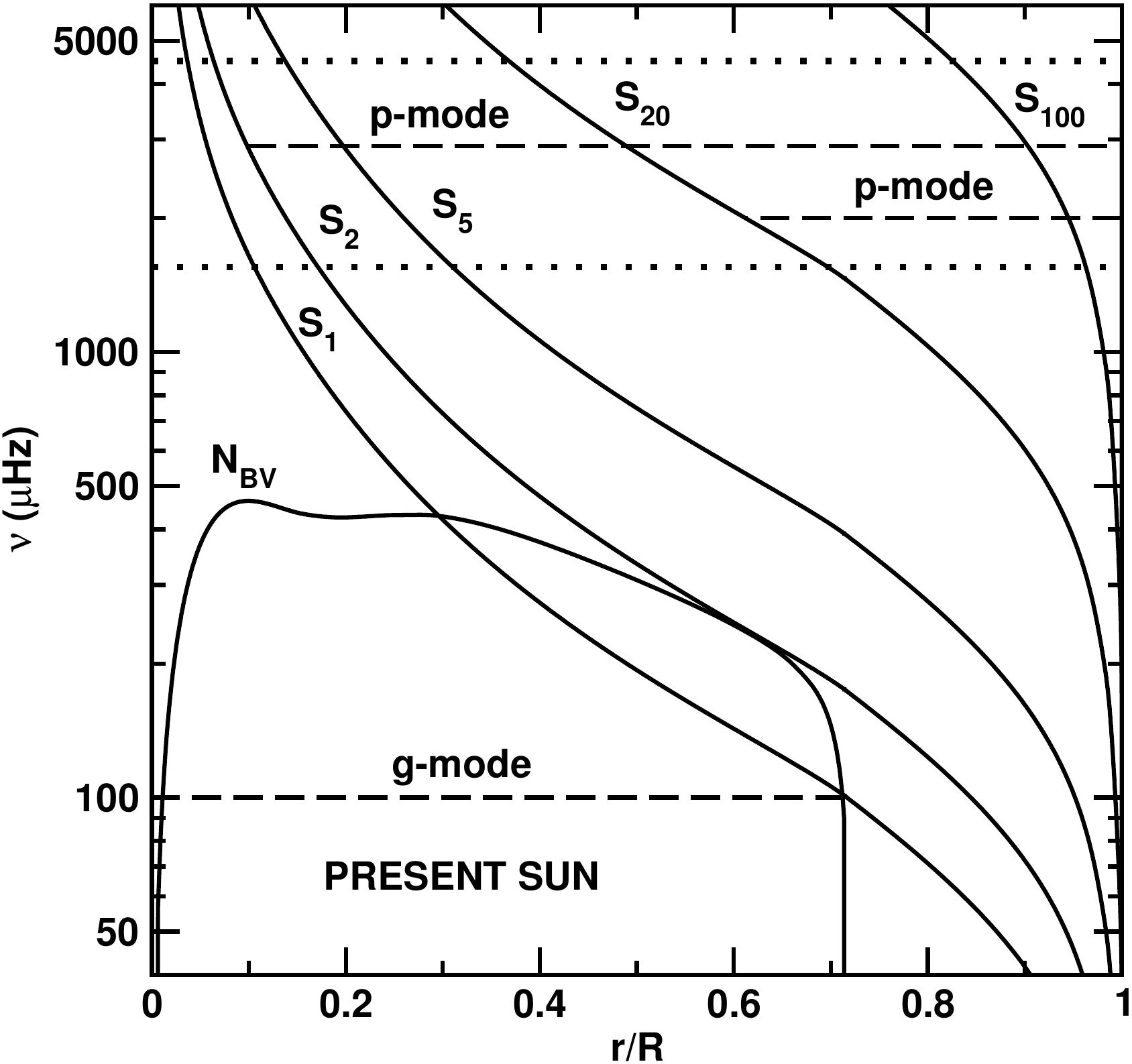}
\caption{Propagation diagram for the Sun. The profiles of the buoyancy $N^2_\mathrm{BV}$ and of the acoustic frequency 
$S^2_{\ell}$, this latter for different values of the angular degree $\ell$, are plotted. Horizontal dotted lines delimit the frequency range 
of observed solar oscillations. From top to bottom, dashed lines span the cavities where $p$-modes with $\ell=20$ and $\ell=2$ are trapped, and the region of propagation of a gravity mode. [From \citet{2009iaus..258..419l}.] 
}
\label{propagSun}
\end{center}
\end{figure}
\begin{figure}
\begin{center}
\includegraphics[width=0.65\textwidth]{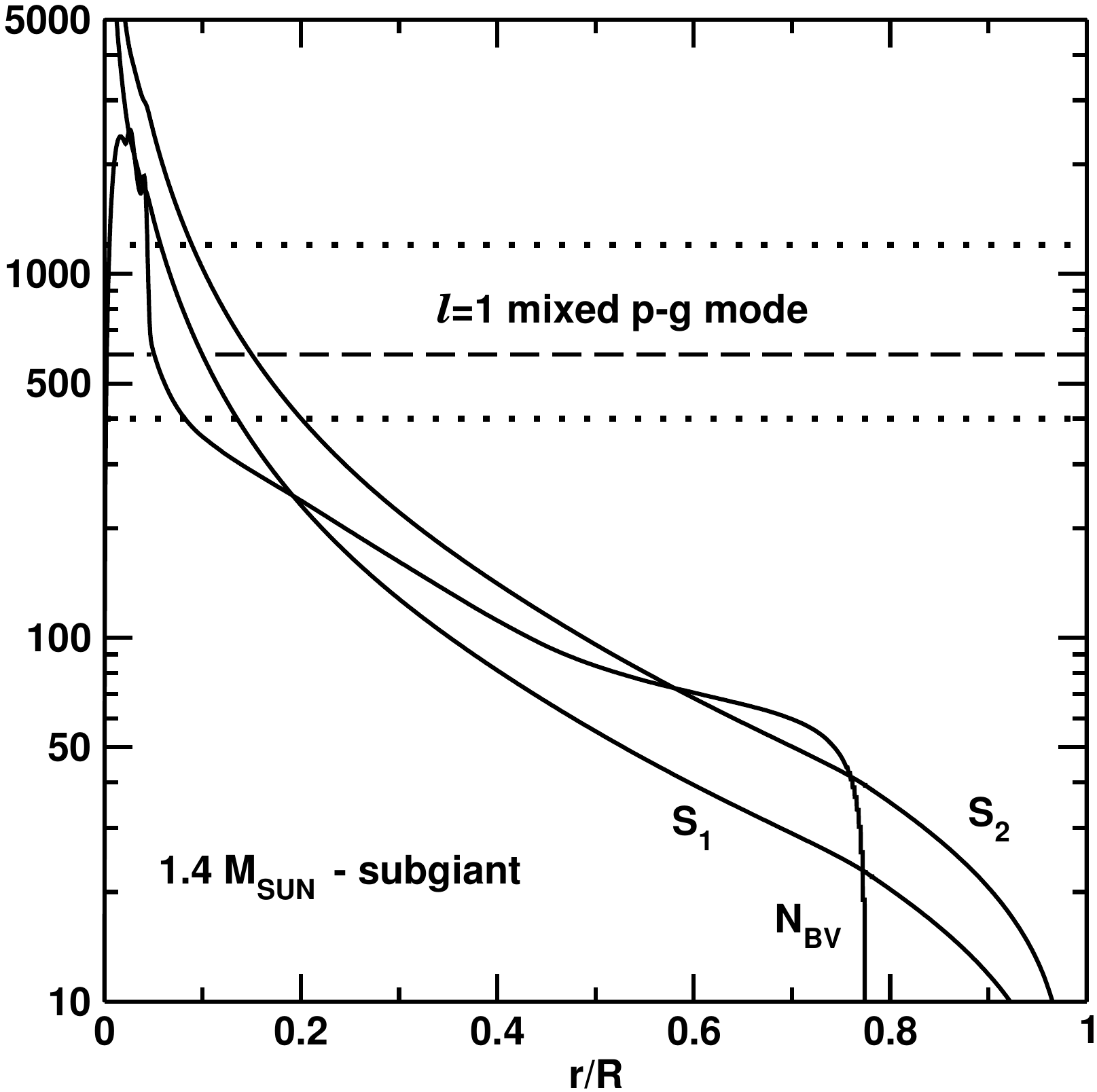}
\includegraphics[width=0.65\textwidth]{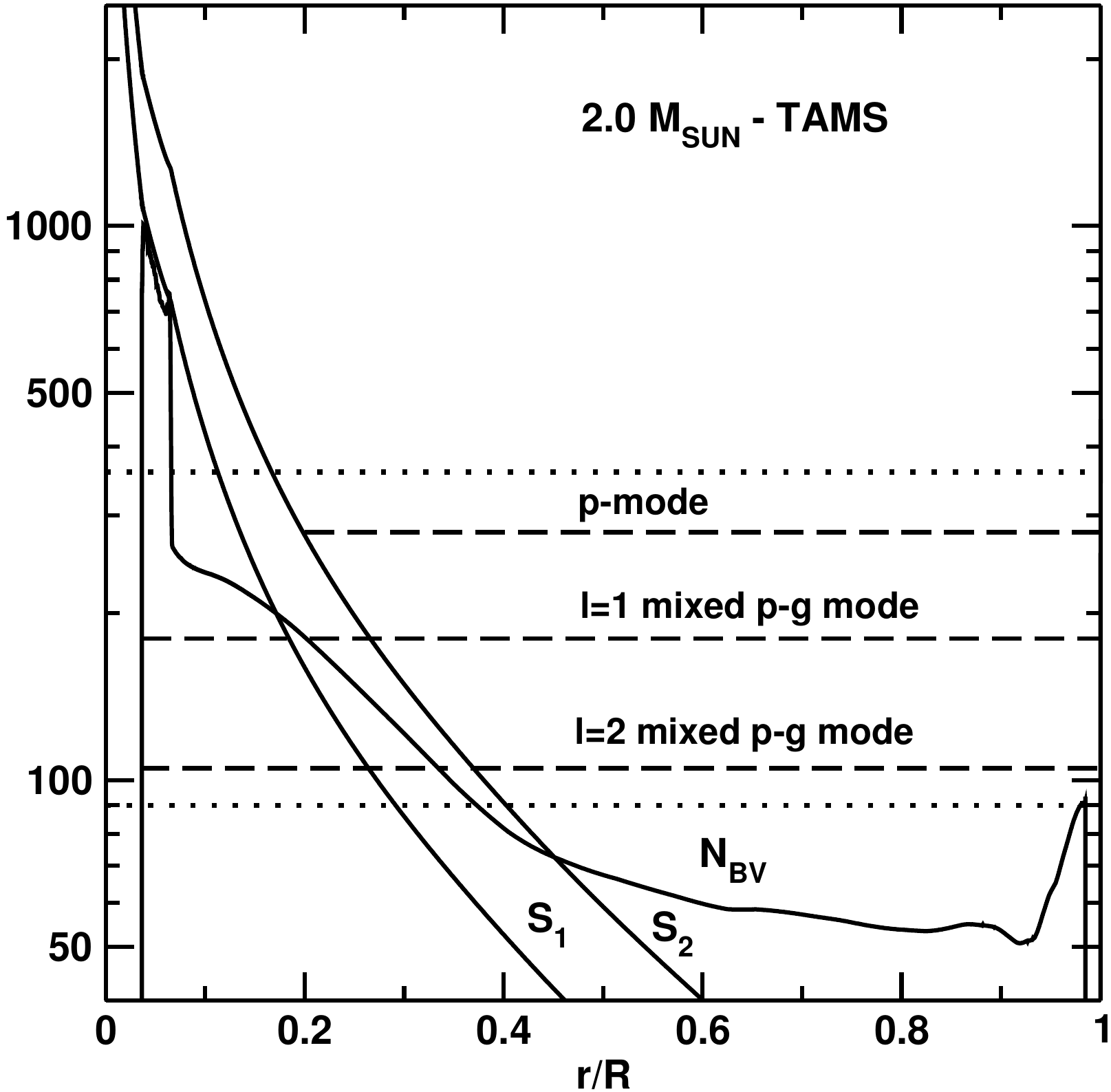}
\caption{Propagation diagrams, as in Fig.~\ref{propagSun} for 
 different stellar masses. 
{\sl Top:}  star of mass $M=1.4\ M_\odot$ in the subgiant stage. 
{\sl Bottom:} star of mass 
$M=2\ M_\odot$ at the end of its MS phase, close to the TAMS. The 
frequency domains (dotted lines) and regions of propagation of $p$-modes 
and mixed $p-g$ modes (dashed lines) are indicated. [From  
\citet{2009iaus..258..419l}.]}
\label{propagStars}
\end{center}
\end{figure}

A further simplification of the problem, called the Cowling approximation \citep{1941MNRAS.101..367C}, consists in neglecting the perturbations of the gravitational potential $\Phi$. For what concerns the present lecture, it is valid at large values of the absolute value of the radial order $n$.
Moreover, some more approximations can be made, which consist in neglecting the terms containing derivatives
of the equilibrium quantities \citep[\eg,][pages 203-204]{2010aste.book.....A}. With these approximations,
the differential equation governing the radial displacement is of second order and reads,
\begin{equation}
\label{eqosc}
\frac{\mathrm{d}^{2} \xi_{r}}{\mathrm{d} r^{2}}+ K_{r} \xi_{r} 
\; = \; 0 \; ,
\end{equation}
with
\begin{equation}
\label{muelle}
K_{r} \; = \; \frac{1}{\omega^2 {c}_\mathrm{s}^2} \left(N^2_\mathrm{BV}
 - \omega^2\right)\left(S^2_{\ell} - \omega^2\right) \; ,
\end{equation}
where $N^2_\mathrm{BV}$ and $ S^2_{\ell} $ are characteristic frequencies defined below.
The nature of the solution depends on the sign of $K_{r}$. When $K_{r}>0$, the solution is an oscillating wave, while when $K_{r}<0$ 
the solution is an exponential function and the wave is evanescent. 
In the adiabatic approximation, the frequencies only depend on two quantities (for instance $c_\mathrm{s}$ and $\rho$).

Two characteristic frequencies appear in Eq.~\ref{eqosc}. The critical acoustic frequency, or Lamb frequency reads,
\begin{equation}
\label{Lamb}
 S^2_{\ell} \; = \; \dfrac{\ell(\ell+1) c_{\mathrm{s}}^2}{r^2} \quad .
\end{equation}

It characterizes the compressibility of the medium. 
Note that $S^2_{\ell} $ depends on the chemical composition inside the star, via the dependence 
in $\mu$ of the sound speed. Therefore, it is related to the evolutionary 
stage and age of the star. The buoyancy $ N^2_\mathrm{BV} $, or Brunt-V\"{a}is\"{a}l\"{a} (BV) 
frequency, reads, 
\begin{equation}
N^2_\mathrm{BV} \; = \; g \left( \frac{1}{\Gamma_1} \, 
 \frac{\mathrm{d} \ln p}{\mathrm{d}r} - \frac{\mathrm{d} 
\ln \rho}{\mathrm{d}r}\right) \; .
\end{equation}
In a radiative zone, $ N^2_\mathrm{BV} >0$.
 It is the frequency associated to the oscillation of a perturbed 
parcel of a gravitationally stratified fluid. 
 In a convective zone $ N^2_\mathrm{BV} <0$.
Furthermore, for an ideal gas,
\begin{equation}
\label{BV}
N^2_\mathrm{BV} \; =  \; \frac{g^2 \rho}{p} \left( \nabla_\mathrm{ad} - 
\nabla + \nabla_\mu\right) \; ,
\end{equation}
where $\nabla=(\mathrm{d}\ln T/ \mathrm{d}\ln p)$ denotes the actual temperature gradient, 
$\nabla_\mathrm{ad}$ the adiabatic gradient, and  $\nabla_\mu=(\mathrm{d}\ln \mu/ \mathrm{d}\ln p)$, 
the gradient of the chemical composition. In a star interior, the profile of $N^2_\mathrm{BV}$ 
reflects density changes, and chemical composition changes via the gradients of the chemical composition built by the evolution.

The solution of Eq.~\ref{eqosc} is an oscillatory one when
\begin{itemize}
\item  $\lvert \omega \rvert > \lvert N_\mathrm{BV} \rvert$ and $\lvert \omega \rvert > \lvert  S_{\ell}\rvert$. In that case, the 
solution corresponds to standing (acoustic) pressure waves, commonly called $p$-modes, 
for which the restoring force is related to the pressure gradient.
\item  $\lvert \omega \rvert < \lvert N_\mathrm{BV} \rvert$ and $\lvert \omega \rvert < \lvert  S_{\ell}\rvert$. 
In that case, the solution corresponds to standing gravity waves, commonly called $g$-modes, for which the restoring force is 
related to the buoyancy.
\end{itemize}

The kind of possible solutions in a star like the Sun are illustrated in the propagation diagram in
 Fig.~\ref{propagSun}. It shows that in the Sun, pressure-modes of $\ell\ge 0$ can propagate up to the surface
where they are observed, while gravity modes are predicted to propagate in the inner regions, but are evanescent 
in the convective envelope. Since the frequencies of the oscillation modes depend on the properties of the stars,
propagation diagrams are different for different stellar masses, evolutionary stages (ages), and chemical
compositions. This is illustrated in Fig.~\ref{propagStars} where we show the propagation diagrams of a
$1.4\ M_\odot$ star in the subgiant stage and of a $2\ M_\odot$ star at the end of its MS phase, close to the
terminal age main sequence (TAMS). This latter would correspond to a A-star in the $ \delta $ Scuti phase. 
We point out that the profiles 
of the buoyancy in these two stars differ from  the solar one. 
 During the MS, those stars possessed a convective core (still present in the 
 $2\ M_\odot$ star). Because of nuclear evolution and convective core recession during the MS, 
 a steep molecular weight gradient has been built in the inner
  regions producing the observed peak in $N_\mathrm{BV}$ (see Eq.~\ref{BV}). 
  Furthermore, the increase of density in the central regions of the $1.4\ M_\odot$ model, 
  resulting from core contraction on the subgiant branch leads to a central increase of $N_\mathrm{BV}$,  and therefore of the frequencies of possible $g$-modes. In these  
  slightly evolved stars, the range of frequencies of gravity modes overlap the one of pressure modes,
  while the evanescent region between the $g$-mode trapping cavity and the $p$-mode one is narrower. 

When the intermediate evanescent region is
narrow enough,  one  $g$-mode and one $p$-mode which happen to have very close frequencies actually 
are  no longer a pure $g$-mode  and a
pure $p$-mode respectively. Instead both modes behave both as a $g$-mode in the inner cavity and as a $p$-mode in the outer cavity.  
This phenomenon called avoided crossing has been predicted and described by \citet{1974A&A....36..107S} and \citet{1977A&A....58...41A}, and 
was first observed by \citet{1975PASJ...27..237O} in a $10\ M_\odot$ star. Such modes, that change their nature when propagating in the star, are 
called mixed  $p-g$ modes. They carry information on the inner regions of the star \citep[see the detailed description of the 
conditions of 
occurrence of these modes by][]{2011a&a...535a..91d}. Mixed  modes provide diagnostics on the compactness of stellar cores as well as on the 
different mixing processes at work during the evolution of a star. This is illustrated in Sect.~\ref{applications}.
 
Fig.~\ref{HRpuls} illustrates the variety of oscillation modes excited in different types of
stars with different internal structures. The main oscillators are listed below.
\begin{itemize}
\item Solar-like oscillators: high radial order $p$-modes with periods spanning the range a few minutes (MS) to a few hours (red giants). 
\item $ \gamma $ Doradus stars: high order $g$-modes, periods in the range $\approx$ 8 hours--3 days.
\item $ \delta $ Scuti stars: low order $p-g$ modes, periods in the range $\approx$ 30 minutes--6 hours. 
\item SPB stars: high order $g$-modes with periods in the range $\approx$ 15 hours--5 days.
\item $ \beta $ Cephei stars: low order $p-g$ modes with periods in the range $\approx$ 2--8 hours. 
\end{itemize}

\begin{figure}
\begin{center}
\includegraphics[width=\textwidth]{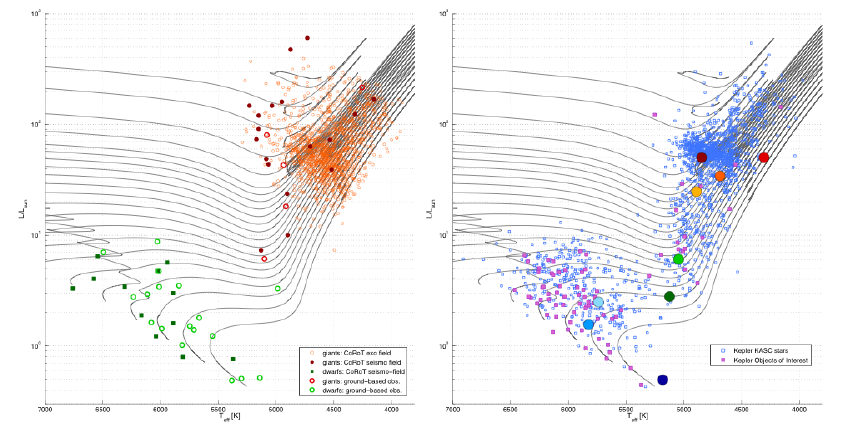}
\caption{Stars where solar-like oscillations are detected, after \citet{2013ARA&A..51..353C}.
{\sl Left:} stars observed during \textsl{CoRoT} long runs, \ie, 3 months of observation per star. 
{\sl Right:}  stars observed by \textsl{Kepler} for a total duration of 3 
years observations per star.}
\label{harvest}
\end{center}
\end{figure}

\subsection{Solar-like oscillators}

The proximity of the Sun permits high temporal and spatial resolution observations. More than $10^7$ 
pressure modes have 
been observed and identified with angular degrees $\ell=0$ to $\sim 5000$. The high number of frequencies and the extremely 
high accuracy of the 
frequency measurements have allowed to conduct sophisticated studies. The inversion of the oscillation frequencies provided the 
profile of the sound speed, of the density, and of the rotation rate from the surface to regions close to the centre (see Sect.~\ref{Sun}). 
From these studies, very accurate 
measurements were drawn, mainly the depth of the convective envelope and its helium content. In the case of stars 
different from the 
Sun, all kinds of modes may propagate depending on the star, that is either $p$, $g$, or $p-g$ mixed modes. However, 
because of the lack of 
spatial resolution and 
of  measurements restricted to integrated light, the accessible modes are those with low 
angular degrees with $\ell$ in the range 0 to 3 at best. Inversions of 
oscillations have been performed for several red giants (RGs) and subgiants (SGs), which provided the rotation profile 
(see Sect.~\ref{redgiants}). A rich harvest of 
observational data 
has been provided by \textsl{CoRoT} and \textsl{Kepler}, as illustrated in Fig.~\ref{harvest}.

\subsection{Asteroseismic analysis and diagnostics}
\label{diagnos}

\begin{figure}
\begin{center}
\includegraphics[width=\textwidth]{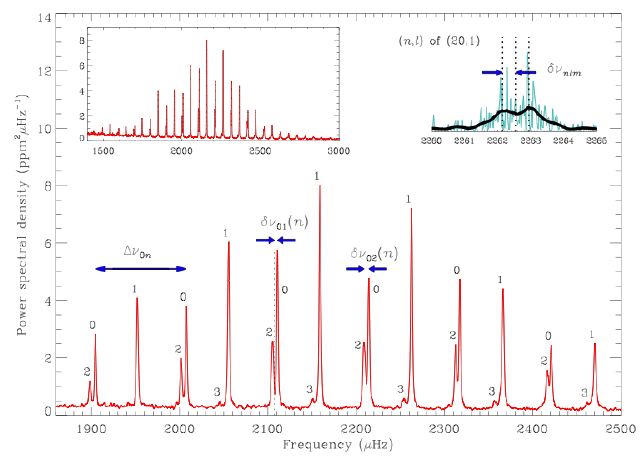}
\caption{ \textsl{Kepler} oscillation spectrum of the MS star 16 Cygni A.
 {\sl Top left-hand inset:} frequency-power spectrum. 
 {\sl Main plot:}  zoom on a range in frequency, indicating the frequency 
of maximum oscillation power, $\nu_\mathrm{max}$ at about 2200 $\mu$Hz,
and the characteristic frequency separations discussed in the text.
{\sl Top right-hand inset: }
zoom in frequency showing rotational frequency splitting of the non-radial 
$\ell = 1$, $n = 20$ mode (not discussed here). 
[From \citet{2013ARA&A..51..353C}.]}
\label{16Cygni}
\end{center}
\end{figure}

\subsubsection{Individual oscillation frequencies}
\label{absfreq}

Figure~\ref{16Cygni}, left-hand inset, from \citet{2013ARA&A..51..353C}, shows the frequency power spectrum of the solar-like oscillator 16 Cygni A, as observed by \textsl{Kepler} \citep{2012ApJ...748L..10M}. 
This power spectrum shows the now famous regular pattern of pressure modes observed in the Sun and in the many solar-like oscillators observed by \textsl{CoRoT} and \textsl{Kepler}. Another example is the frequency power spectrum of the 
 \textsl{CoRoT} main target HD~43587 obtained by \citet{2014A&A...564A..34B}.

A formulation adapted from the asymptotic expansion by \citet[][]{1967AZh....44..786V} and \citet[][]{1980apjs...43..469t} is commonly used to  interpret the 
observed low-degree pressure mode oscillation spectra \citep[see \eg,][and references therein]{2013A&A...550A.126M}. It approximates 
the frequency of a $p$-mode of high radial order $n$ and angular degree $\ell{\ll}n$ as
\begin{equation}
\label{asym}
\nu_{n,\ell} \; {\simeq} \; \left( n+\frac{1}{2}\ell
 +\epsilon\right)\Delta \nu_\mathrm{asym}-\ell(\ell+1)D_0 \; .
\end{equation}
Therefore, in the asymptotic approximation, $p$-modes of same angular degree and consecutive $n$ are nearly equidistant in frequency.
The coefficients $\epsilon$, $\Delta \nu_\mathrm{asym}$, and $D_0$ depend on the considered equilibrium state of the star.
In particular
\begin{equation}
\label{asymstuff}
\Delta \nu_\mathrm{asym} \; {=} \; \left( 2\int\limits_{0}^{R} 
\frac{\mathrm{d}r}{c} \right)^{-1} \ \mathrm{and}\ \ 
D_0\approx -\frac{\Delta \nu_\mathrm{asym}}{4\pi^2 \, 
\nu_{n,\ell}} \int\limits_{0}^{R} \frac{\mathrm{d}c}{r} \; .
\end{equation}
 
 The quantity $\Delta \nu_\mathrm{asym}$ measures the inverse of the sound 
 travel time across a stellar diameter and is proportional to the square root of the mean density.
 The quantity $D_0$
 probes the evolution status (and then age of the star) through the sound speed gradient built by the
 chemical composition changes in the stellar core.
 
 The $\epsilon$ term weakly depends on $n$ and $\ell$, but it is much sensitive to the physics of surface layers. 
 The problem is that outer layers in solar-type oscillators are the seat of inefficient convection, a 3-D,  
 non-adiabatic, and turbulent process, which is poorly understood. 
 The modelling of near surface stellar layers is uncertain and so are the related computed frequencies. 
These so-called near-surface effects are a main concern when using individual frequencies to constrain 
stellar models because they are at the origin of an offset between observed and computed oscillation frequencies. 
Some empirical recipes are used to correct for this offset. This is discussed below. 
 
 \subsubsection{$\nu_\mathrm{max}$, $\langle \Delta \nu \rangle$ and scaling relations}
 \label{scaling}
  
 From the oscillation power spectrum, the frequency at maximum amplitude $\nu_\mathrm{max}$ 
is extracted (see left-hand inset in Fig.~\ref{16Cygni} where $\nu_\mathrm{max, 16 Cyg}\simeq 2200~\mu\mathrm{Hz}$). 
 This quantity is proportional to the acoustic cut-off frequency, itself related to effective temperature 
 and surface gravity \citep[see \eg,][]{1994ApJ...427.1013B, 1995A&A...293...87K,2011A&A...530A.142B}. 
 This dependency yields a scaling relation used to constrain the mass and radius (or surface gravity) of a star of known $T_\mathrm{eff}$,
 \begin{equation}
 \label{scalingnumax}
 \frac{\nu_\mathrm{max, sc}}{\nu_\mathrm{max,\odot}} \; {=} \; 
\left(\frac{M}{M_\odot}\right)
 \left(\frac{T_\mathrm{eff}}{5777 \, K}\right)^{-1/2} 
\left(\frac{R}{R_\odot}\right)^{-2} \, = \, 
\left(\frac{g}{g_\odot} \right)
 \left(\frac{T_\mathrm{eff}}{5777\,K}\right)^{-1/2} \; ,
 \end{equation}
  where  $\nu_\mathrm{max,\odot}=3050\ \mu$Hz is the solar value and the index $\mathrm{sc}$ stands for scaling.
 
 The difference in frequency of two $p$-modes of same degree and consecutive orders reads
 \begin{equation}
 \label{indivsep}
 \Delta \nu_{\ell}(n) \; {=} \; \nu_{n,\ell}{-}\nu_{n-1,\ell} \, , 
 \end{equation}
 and is named the large frequency separation (see main panel of Fig.~\ref{16Cygni}). 
 The mean large frequency separation is used to constrain stellar models.

There are different ways to estimate the large frequency separation in a stellar model. First, it can be calculated as an average of the individual separations defined by Eq.~\ref{indivsep}. Second, in the asymptotic approximation (Eq.~\ref{asym}), $\Delta \nu_{\ell}(n){\equiv}\Delta \nu_\mathrm{asym}$ is approximately constant whatever the $\ell$ value.  $\Delta \nu_\mathrm{asym}$ can therefore be derived from an adjustment of the asymptotic relation through the observed frequencies (for instance by a weighted least-squares fit). Such an adjustment also provides the values of $D_0$ and $\epsilon$.
 Third,  as mentioned above,  $\langle \Delta \nu \rangle \propto \langle \rho\rangle^{1/2}$.
 This yields a scaling relation usable to constrain stellar mass and radius \citep[see \eg,][]{1986apj...306l..37u, 1995A&A...293...87K},
 \begin{equation}
 \label{scalingdeltnu}
 \frac{\langle \Delta \nu \rangle_\mathrm{sc}}{\langle \Delta \nu \rangle_\odot} {=} \left(\frac{M}{M_\odot}\right)^{1/2} \left(\frac{R}{R_\odot}\right)^{-3/2},
 \end{equation}
 where $\langle \Delta \nu \rangle_\odot=134.9\ \mu$Hz is the solar value.
  
   \subsubsection{Small frequency separations and separation ratios}
 \label{separ}  
 
 The difference in frequency of two pressure modes of degrees differing by two units and orders differing by one unit reads
 \begin{equation}
 d_{\ell,\ \ell+2}(n) \; {=} \; \nu_{n,\ell}{-}\nu_{n-1,\ \ell+2} \quad , 
 \end{equation}
 and is commonly referred to as the small frequency separation. 
  According to the asymptotic relation (Eq.~\ref{asym}), $d_{02}(n)$ scales as $ {\approx} 6 D_0$. 
  Thus, $d_{02}(n)$ probes the central conditions, and in turn the evolutionary status of stars.
  The quantity $d_{02}(n)$, often also denoted by $\delta\nu_{02}(n)$, is shown in Fig.~\ref{16Cygni}.
  
 Modes of $\ell=1$  are rather easy to detect, while  modes of $\ell=2$ are not always observed, or are affected by large error bars.
 This led \citet{2003A&A...411..215R} to propose to use either the three points small frequency 
 separations $d_{01}(n)$ and $d_{10}(n)$, or the five points separations 
 $dd_{01}(n)$ and $dd_{10}(n)$ as diagnostics for stellar models. These quantities read
 \begin{eqnarray}
 d_{01}(n) &  = & -\frac{1}{2}(\nu_{n-1, 1}-2\nu_{n, 0}+\nu_{n, 1}) \, \\ 
 d_{10}(n) & = & \frac{1}{2}(\nu_{n+1, 0}-2\nu_{n, 1}+\nu_{n, 0}) 
 \end{eqnarray}
 for the three points separations, and 
 \begin{eqnarray}
 dd_{01}(n) & =  & \frac{1}{8}(\nu_{n-1, 0}-4\nu_{n-1, 1}+
6\nu_{n, 0}-4\nu_{n, 1}+\nu_{n+1, 0}) \, , \\
 dd_{10}(n) & = & -\frac{1}{8}(\nu_{n-1, 1}-4\nu_{n, 0}
+6\nu_{n, l}-4\nu_{n+1, 0}+\nu_{n+1, 1}) \, ,
 \end{eqnarray}
  for the five points separations.
An illustration is provided in Fig.~\ref{16Cygni} for the quantity $d_{01}(n)$ denoted by $\delta\nu_{01}(n)$.
According to the asymptotic relation (Eq.~\ref{asym}), $dd_{01/10}(n)$ scales as $ {\approx} 2 D_0$. Thus, these
quantities also probe the central
conditions, and in turn the evolutionary status of stars.
 
 Furthermore, \citet{2003A&A...411..215R} demonstrated that, while the frequency separations are sensitive to near-surface effects, these effects nearly cancel in the frequency separation ratios defined as
 \begin{eqnarray}
 r_{02}(n) &  = & d_{02}(n)/\Delta \nu_{1}(n) \; , \\
 rr_{01}(n) & = & dd_{01}(n)/\Delta \nu_{1}(n) \; , \\ 
 rr_{10}(n) &  = & dd_{10}(n)/\Delta \nu_{0}(n+1) \; .
 \end{eqnarray}
 
Furthermore,  the second difference $\delta_2 \nu$ has been 
introduced by \citet{1990LNP...367..283G}:
 \begin{equation}
 \delta_2 \nu_{n,\ell} \; = \; \nu_{n+1,\ell}-2\nu_{n,\ell}
+\nu_{n-1,\ell} \; , 
 \end{equation}
 and can be seen as a second derivative of the frequency $\nu_{n,\ell}$ with respect to $n$.
This quantity is sensitive to rapid variations in the sound speed profile inside the star.
Frequency separations and ratios can be used to constrain stellar models.
 
\subsubsection{Period separations for $g$-dominated mixed modes}
\label{period-spacing}

  In the asymptotic theory \citep[][]{1980apjs...43..469t}, it is shown that the period difference between two 
  pure $g$-modes of same degree and consecutive radial orders is approximately constant and reads,
  \begin{equation}
  \label{pi_spacing}
   \Delta P_{n,\ell} \;  = \;P_{n+1,\ell}-P_{n,\ell} \; \propto \; 
\frac{n}{\sqrt{\ell(\ell+1)}}\Pi_0 \; ,
   \end{equation}
   where
 \begin{equation}
 \Pi_0 \; {=} \; \left(\ \int\limits_{r_\mathrm{cc}}^{R} \frac{\lvert
     N_\mathrm{BV} \rvert}{r} dr \right)^{-1} \; .
 \label{pi0}
 \end{equation}  
As discussed above, mixed $p-g$ modes develop in evolved solar-like stars reaching the end of the MS and beyond.
The asymptotic approximation is no longer valid for mixed modes, 
nor for modes whose wavelength is longer than the length-scale 
of variations of the structure of the star in the region where they propagate.
Therefore, the periods of mixed modes described in Sect.~\ref{pgmode},
or those of high-order  $g$-modes propagating in a region with sharp features in $N_\mathrm{BV}$, may deviate with respect to the 
equidistant behaviour predicted by Eq.~\ref{pi_spacing}.
The properties of this deviation allow us to probe the size of the convective core, mixed regions, and extra-mixing processes inside the 
star, as shown by \citet{2007MNRAS.377..373M}, \citet{2008MNRAS.386.1487M}, \citet{2011a&a...535a..91d}, see also 
Sect.~\ref{applications}.

 \subsubsection{Related seismic diagnostics}
 \label{diagnostics}
 
 \begin{figure}
 \begin{center}
\includegraphics[width=0.7\textwidth]{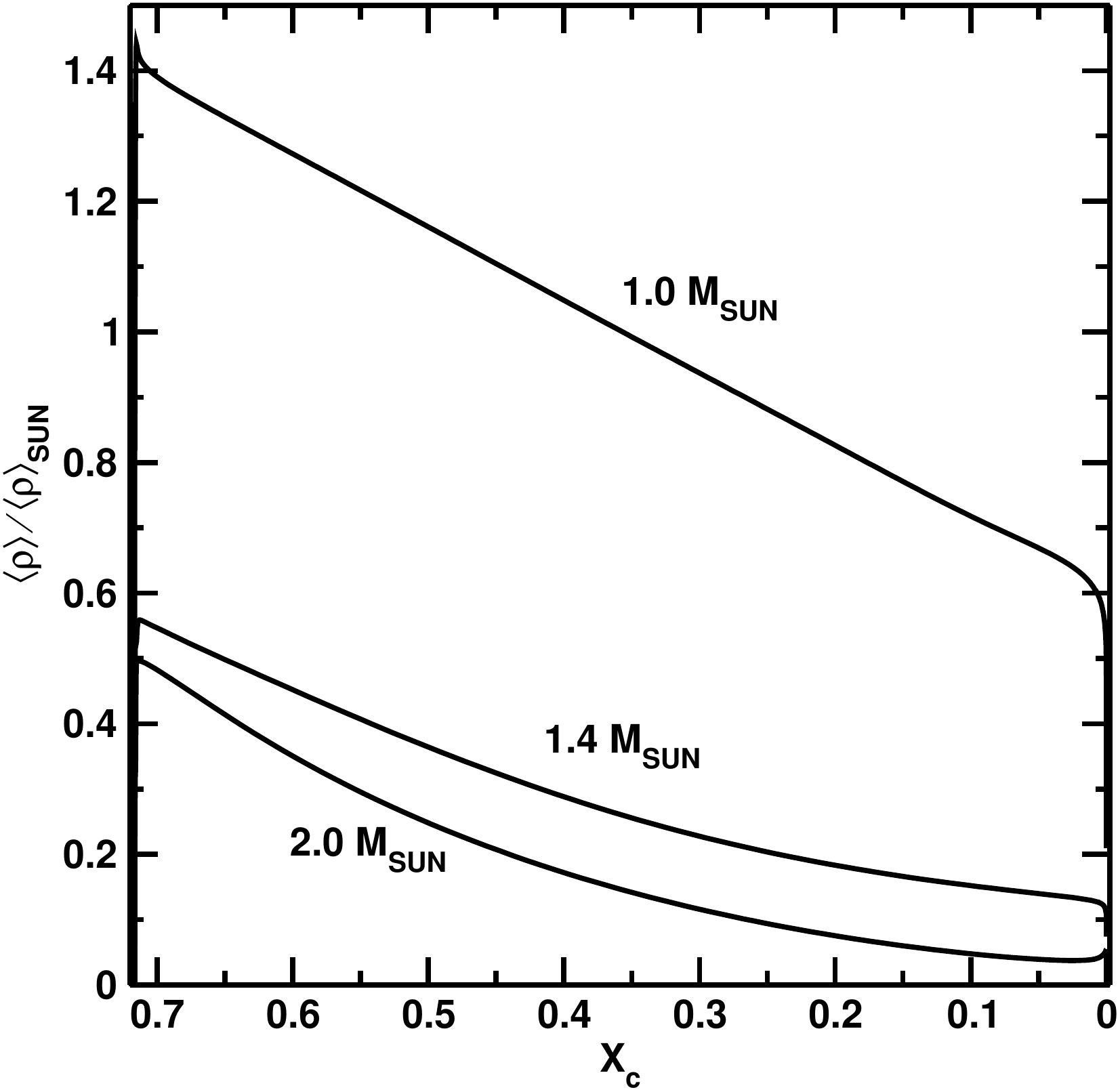}
\includegraphics[width=0.7\textwidth]{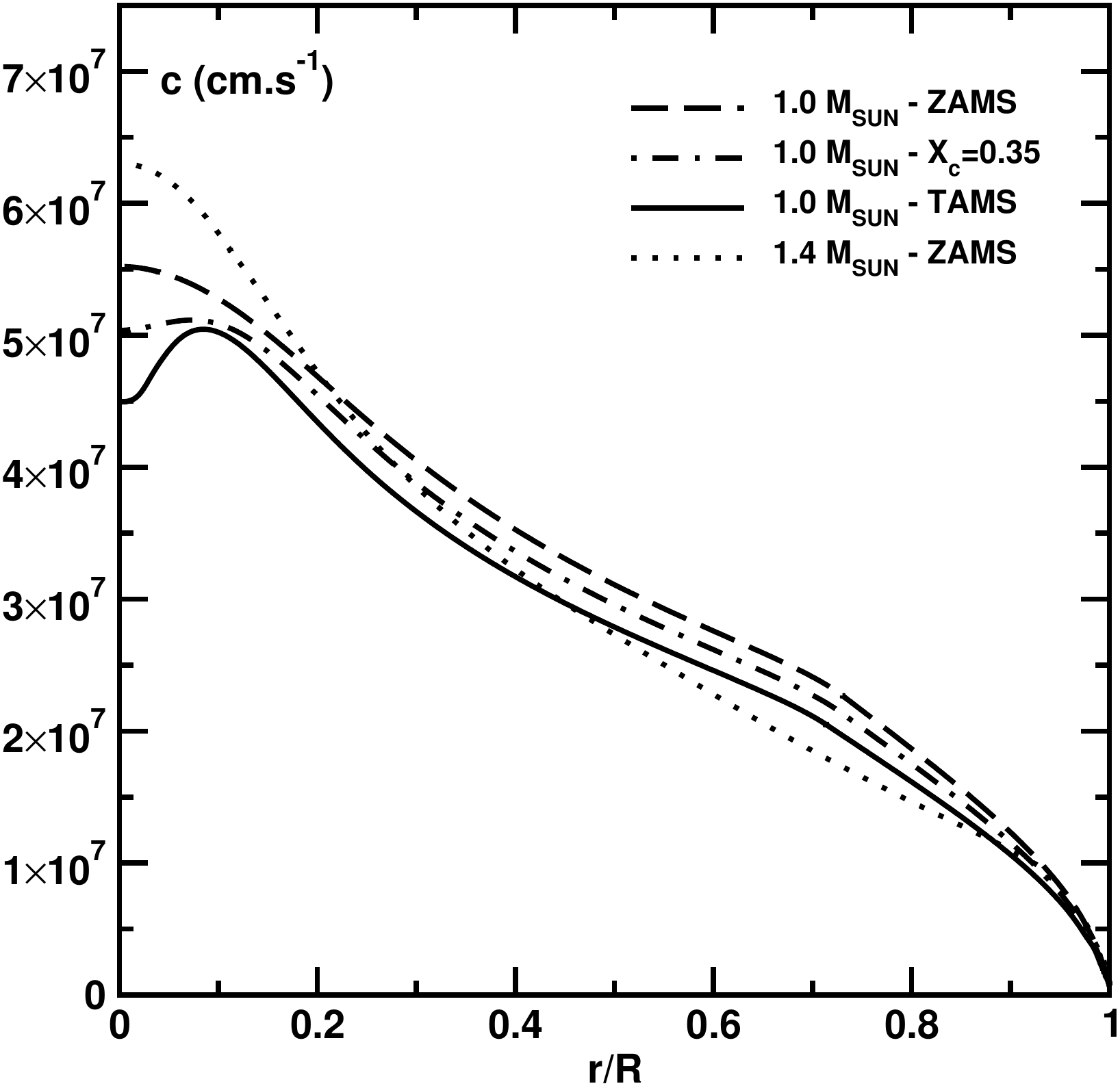}
  \caption{{\sl Top:} the change of the mean density with evolution 
on the MS for three values of the stellar mass. {\sl Bottom:} 
the sound speed profile in $1\ M_\odot$ and $1.4\ M_\odot$
 models, on the ZAMS, in the middle of the MS, and at the end 
of the MS (TAMS).}
 \label{rhosound}
  \end{center}
 \end{figure}
 
The top panel of Fig.~\ref{rhosound} shows the variation during the MS of the mean density $\langle \rho \rangle$ 
in stellar models 
of different masses. It shows that at a given evolutionary stage (\ie, for given central hydrogen mass fraction
 $X_\mathrm{c}$) the more massive the star, the lower $\langle \rho \rangle$, and the more evolved the star, the lower 
 $\langle \rho \rangle$. Since the mean frequency separation of $p$-modes scales as the square root of $\langle \rho \rangle$, at 
 a given evolutionary stage on the MS, one predicts that $\langle \Delta \nu \rangle$ decreases when mass increases. Also, for a 
 given mass, $\langle \Delta \nu \rangle$ is predicted to decrease as evolution proceeds on the MS. 

Figure~\ref{rhosound}, bottom panel, shows the inner sound speed profile in stellar models of different mass at different 
evolutionary stages. It shows that on the ZAMS, the sound speed regularly decreases from the centre to the surface and 
that the higher the mass, the higher the absolute value of the sound speed gradient. Therefore, one predicts that on the ZAMS, the 
small frequency separation $d_{02}$, which is proportional to $D_0$ (Eq.~\ref{asymstuff}) is slightly larger at higher mass. 
On the other hand, for a given mass, the central decrease of the sound speed resulting from the decrease of $X_\mathrm{c}$ (and 
increase of $\mu$) creates a positive sound speed gradient in the centre and therefore a negative contribution to $D_0$; the 
more evolved the star, the higher the central gradient. Thus, one predicts that $d_{02}$ decreases as 
evolution proceeds on the MS. The strong sensitivity of $D_0$ to the core structure makes the small frequency separation a very 
sensitive age indicator. In the future, this behaviour might be used to derive a scaling relation between the small frequency 
separation $d_{02}$ and stellar parameters (mass, evolutionary stage, etc.).

 \begin{figure} 
 \begin{center}
\includegraphics[width=0.7\textwidth]{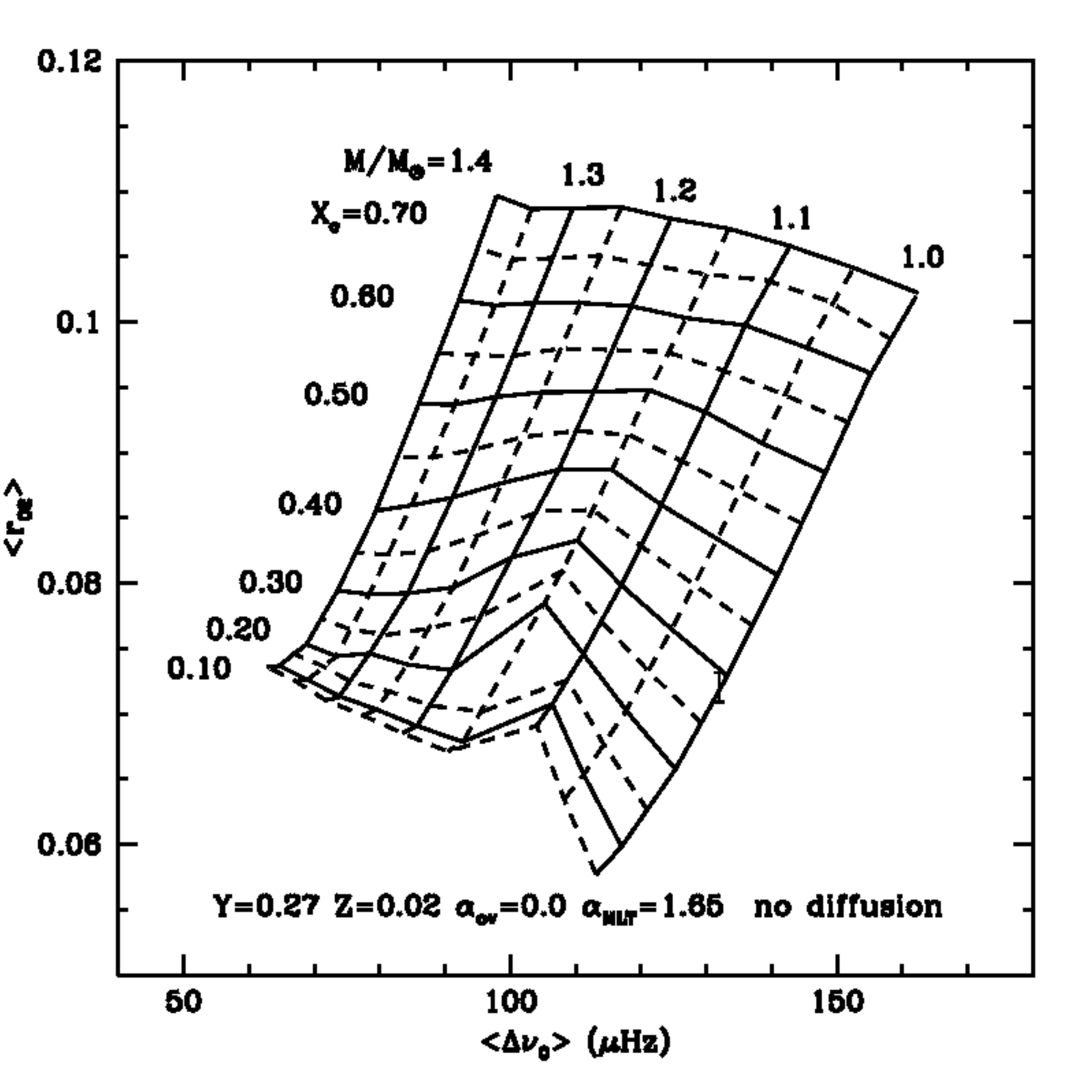}
\includegraphics[width=0.7\textwidth]{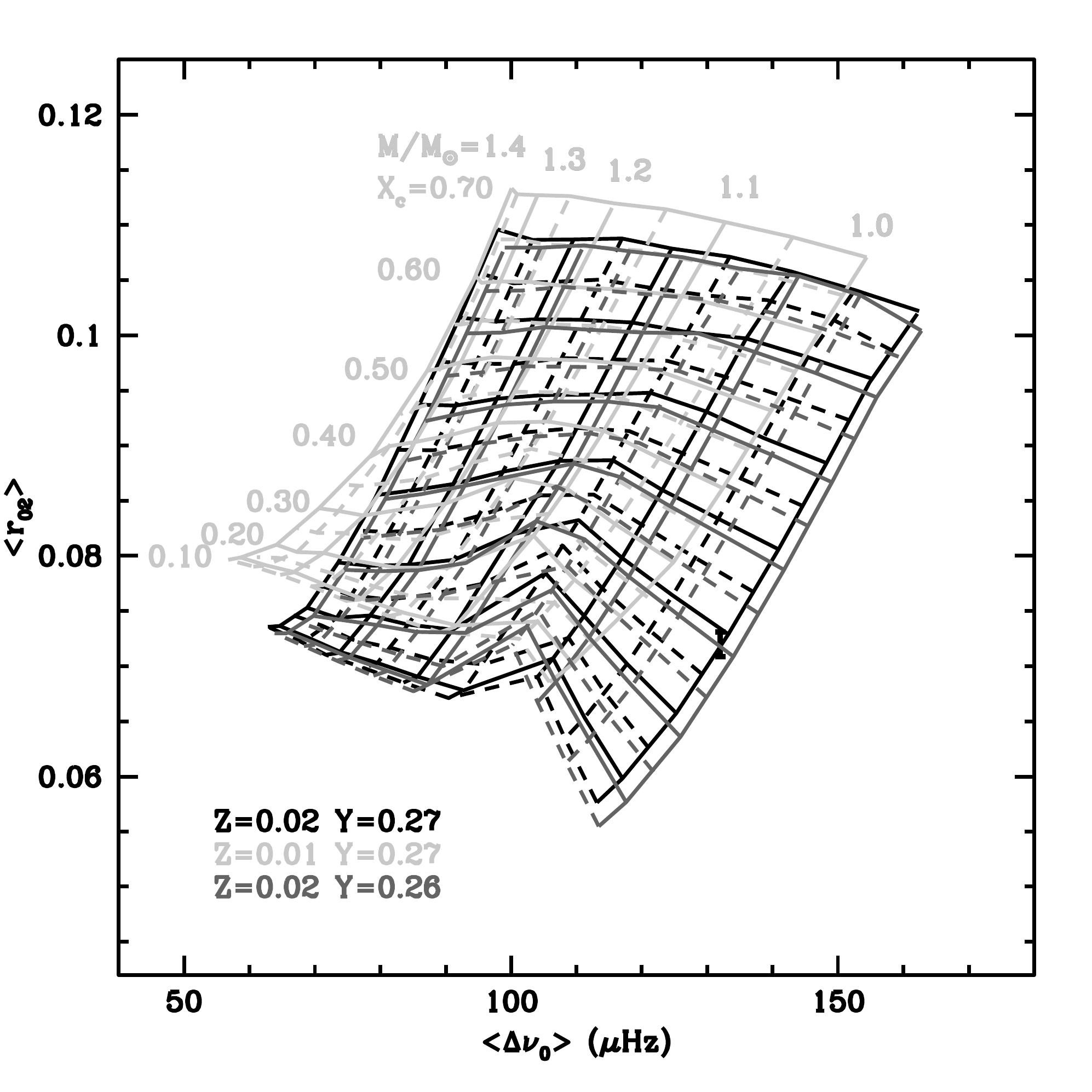}
 \caption{{\sl Top:} : asteroseismic diagram showing the run of $\langle r_\mathrm{{02}}\rangle$ as a function of 
$\langle \Delta \nu\rangle$ for stars of masses in the range $1.0-1.4\ M_\odot$ during the MS. 
Models have been calculated with an initial helium abundance $Y=0.27$, metallicity $Z=0.02$ and mixing-length parameter 
$\alpha_\mathrm{conv}=1.60$ without microscopic diffusion
 (see Lecture 1). 
Evolutionary stages with decreasing central hydrogen abundance $X_\mathrm{c}$ are indicated. 
 {\sl Bottom:} the same, but for different chemical composition sets.}
  \label{r02A}
   \end{center}
 \end{figure}

  \begin{figure} 
  \begin{center}
\includegraphics[width=0.7\textwidth]{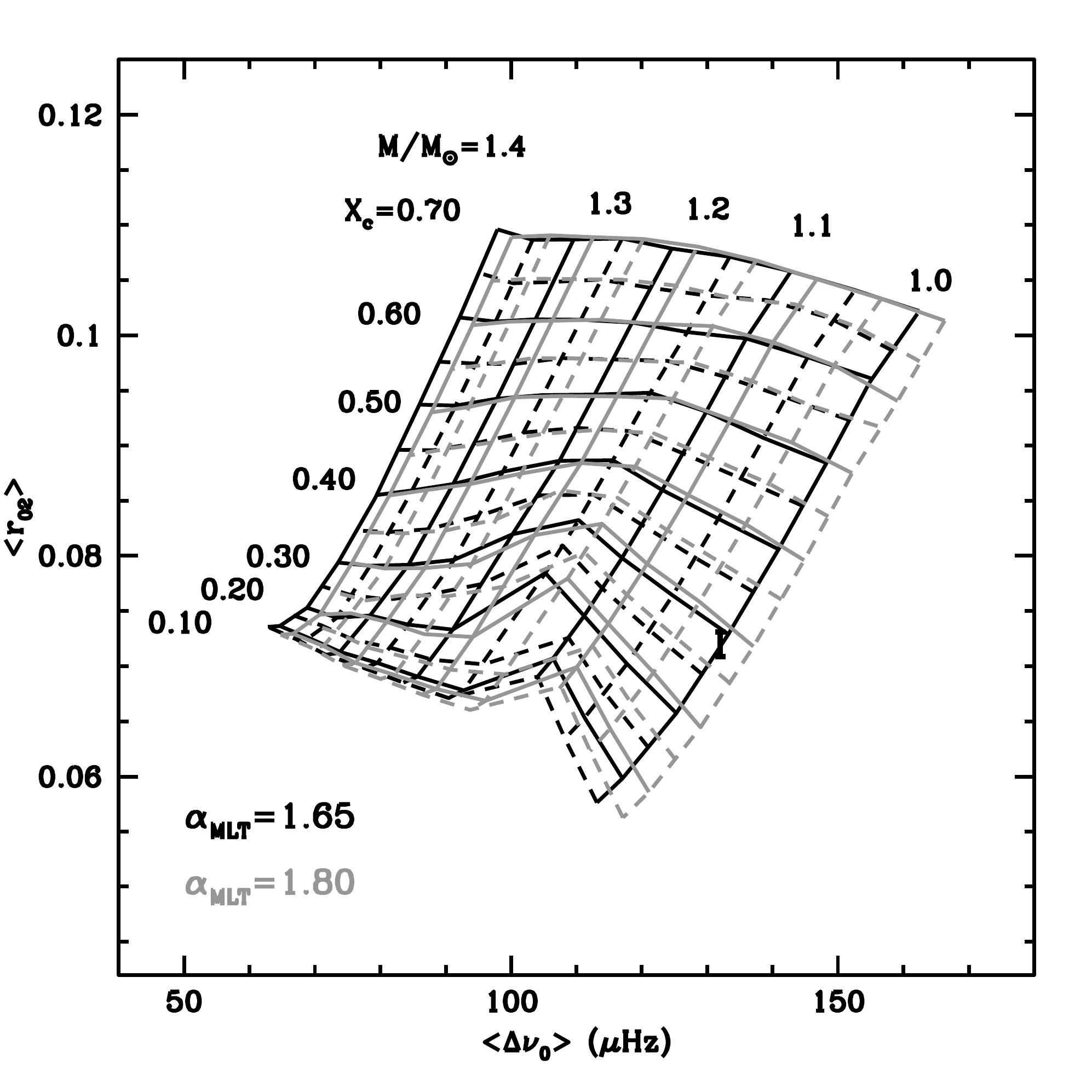}
\includegraphics[width=0.7\textwidth]{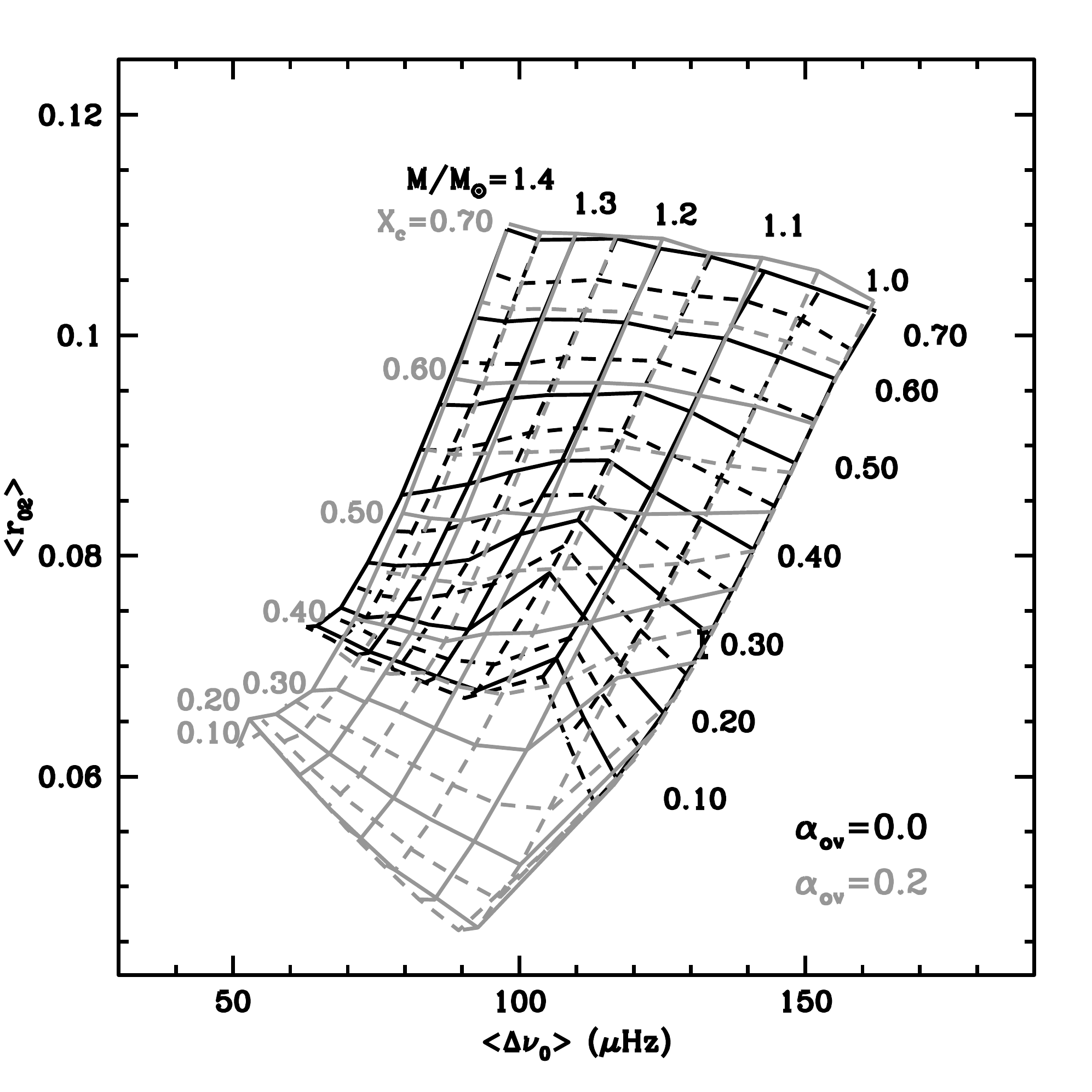}
  \caption{{\sl Top:} the same as in Fig.~\ref{r02A}, but for models with 
two different values of the mixing-length parameter $\alpha_\mathrm{conv}=1.60$ and $1.85$. 
{\sl Bottom:} same diagram for models including overshooting of 
the convective core.
  }
     \label{r02B}
  \end{center}
  \end{figure}
 
\citet{1986apj...306l..37u} and \citet{1988IAUS..123..295C} proposed to use the pair
 ($\langle\Delta\nu\rangle$, $\langle d_{02}\rangle$) as a diagnostic of age and mass of MS stars. 
 To minimize near-surface physics the ($\langle\Delta\nu\rangle$,
 $\langle r_{02}\rangle$) pair can be used instead \citep{2005MNRAS.356..671O}.  Figure~\ref{r02A}, top panel, shows the 
 variations of $\langle\Delta\nu\rangle$ and $\langle r_{02}\rangle$ on the MS, for various masses, at different evolutionary 
 stages. It shows that indeed, on the MS, $\langle\Delta\nu\rangle$ can serve as a mass diagnostic, while $\langle r_{02}\rangle$ 
 is a good evolutionary stage diagnostic. For instance, in the case of the Sun, $\langle\Delta\nu\rangle= 134.9\ 
 \mu\mathrm{Hz}$ and $\langle r_{02}\rangle\sim 0.074$ which by direct reading on Fig.~\ref{r02A} corresponds to a star of 
 about $1\ M_\odot$, with $X_\mathrm{c}\sim 0.30$, while for the \textsl{CoRoT} main target HD~52265 
 ($\langle\Delta\nu\rangle= 98.3\pm 0.1 \, \mu\mathrm{Hz}$ and $\langle r_{02}\rangle\sim 0.084\pm 0.003$), the mass is around $1.2\ M_\odot$ and $X_\mathrm{c}\sim 0.35$. 
 
 On the basis of the precision on frequency measurements performed with \textsl{CoRoT} and \textsl{Kepler} 
 (\ie, an absolute precision of $\sim 10^{-4}$ on individual frequencies), 
 one can predict that the mass and the age of a solar-like oscillator are measured with a precision better than 
 $5$ and $10$ per cent respectively. However, this estimate is made for a given set of stellar models, 
 for chosen input physics and free parameters. As shown in Fig.~\ref{r02A}, bottom panel and Fig.~\ref{r02B}, 
 the age and mass estimates may be hampered if the metallicity of the star is imprecise, 
 or if the initial helium content assumed in the models does not correspond to the actual helium content 
 of the star because the ($\langle\Delta\nu\rangle$, $\langle r_{02}\rangle$) grids are stretched and shifted 
 by composition changes. This calls for precise metallicity determinations in stars 
 \citep[see also Fig.~2 in][]{2013apj...769..141s}.
Also, if the input physics are not appropriate, the ($\langle\Delta\nu\rangle$, $\langle d_{02}\rangle$) 
diagram may be severely distorted (see the strong impact of overshooting in Fig.~\ref{r02B}, bottom panel). 
Various morphologies of the diagram and different ages would result from a bad evaluation of the chemical 
composition gradient in the central regions of stars resulting from overshooting, microscopic diffusion, 
or rotation-induce mixing, etc. Therefore, 
Figs.~\ref{r02A} and ~\ref{r02B} indicate that to improve the age-dating of ensembles of stars, 
it is mandatory to better characterize the physical processes governing their structure and to have a 
more precise estimate of their helium content. This implies to fully calibrate well-known stars (calibrators).
 
\begin{figure}[!ht]
 \begin{center}
\includegraphics[width=0.9\textwidth]{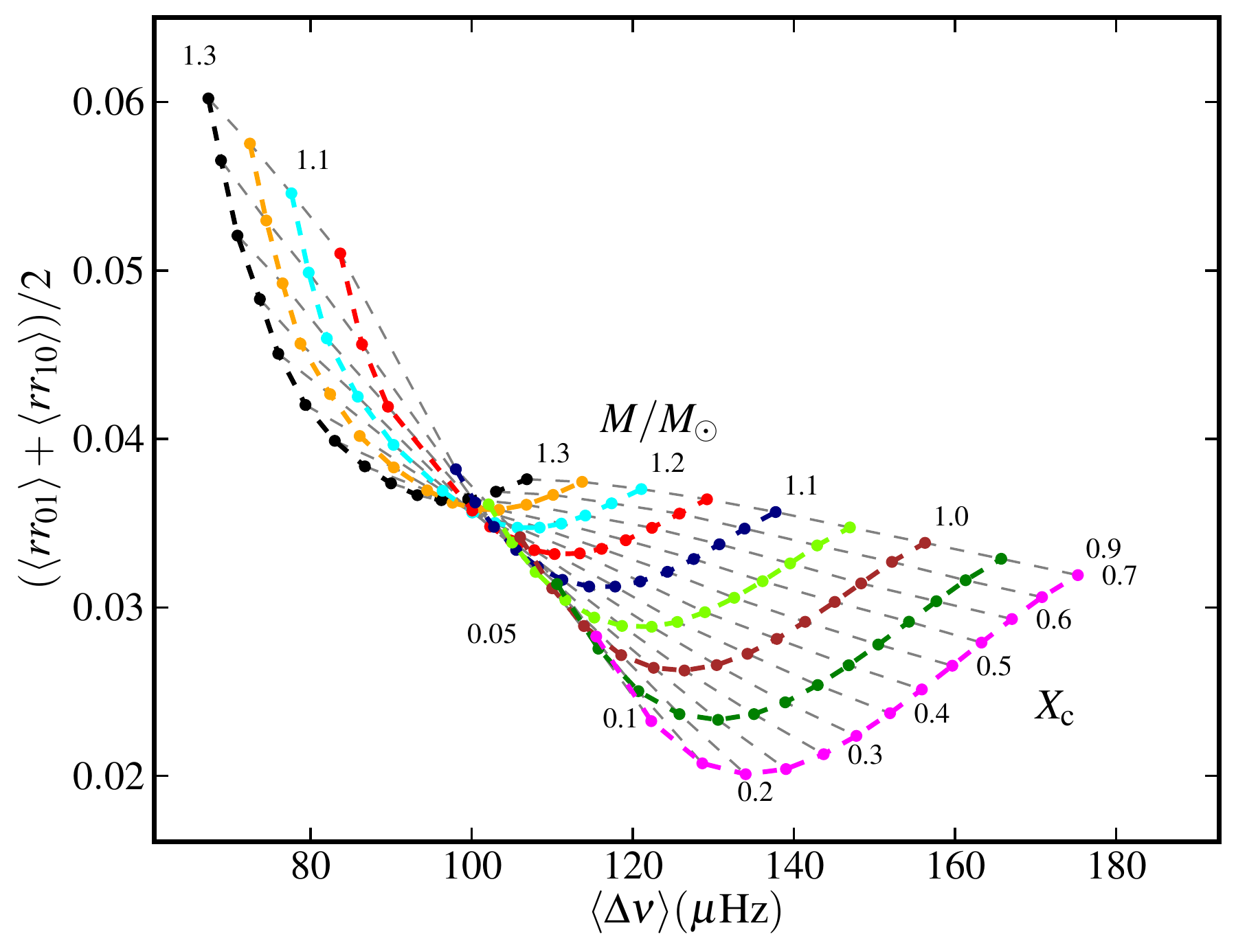}
 \caption{Asteroseismic diagram showing the run of $(\langle rr_\mathrm{{01}}\rangle+ \langle rr_\mathrm{{10}}\rangle)/2$ 
as a function of $\langle \Delta \nu\rangle$ for stars of masses in the range $0.9-1.3\ M_\odot$ in the MS. 
Models have been calculated for an initial helium abundance $Y=0.275$, metallicity $Z/X=0.0245$ and mixing-length value 
$\alpha_\mathrm{conv, CGM}=0.60$ (see Lecture 1). 
 Evolutionary stages with decreasing central hydrogen
 abundance $X_\mathrm{c}$ are pinpointed. 
[From \citet{2014arXiv1406.0652L}.] 
 }
   \label{astdiag}
 \end{center}
 \end{figure}

 The advantage of $\langle r_{02}\rangle$ is that it decreases regularly as evolution proceeds on the MS. 
 But, when only modes of $\ell=1$ are observed, it is interesting to consider the mean ratios 
 $\langle r_{01/10}\rangle$, which are also sensitive to age
 \citep[see \eg,][]{2005a&a...441.1079m,2005a&a...441..615m,2014arXiv1406.0652L}.
   This is illustrated in Fig.~\ref{astdiag} that shows the run of 
   $(\langle rr_\mathrm{{01}}\rangle+ \langle rr_\mathrm{{10}}\rangle)/2$ 
   as a function of $\langle \Delta \nu\rangle$ along the evolution on the 
   MS of stars of different masses. For all masses, the $\langle rr_{01/10} \rangle$ ratio 
   decreases at the beginning of the evolution on the MS down to a minimum and then increases up to the end of the MS. 
   The minimum value is larger and occurs earlier as the mass of the star increases, 
   \ie, as a convective core appears and develops.
 
 Deviations from the asymptotic theory are found in stars, as soon as steep gradients of physical quantities are built. This occurs for instance at the boundaries of convective zones due to the abrupt change of energy transport regime. Such glitches impact the sound speed, and an oscillatory behaviour is then visible in frequency differences \citep[see \eg, ][]{1990LNP...367..283G, 1994A&A...282...73A, 1994MNRAS.268..880R}.  For instance 
 \citet{2012A&A...544L..13L} used the oscillatory behaviour of the observed $rr_{01}(n)$ 
 and $rr_{10}(n)$ ratios to estimate the amount of penetrative convection below the convective envelope of HD~52265.
 
  \subsubsection{Near-surface effects }

As mentioned in Sect.~\ref{absfreq}, near-surface effects are at the origin of an offset between
 observed and computed oscillation 
frequencies  \citep{1997MNRAS.284..527C}. 
This is a major caveat when trying to match model frequencies to observations.
 To cope with this problem, \citet{2003A&A...411..215R} suggested to use instead the frequency separation ratios as 
 stellar models constraints. This approach has been followed by \eg, \citet{2005a&a...441..615m}, 
 \citet{2013apj...769..141s}, and \citet{2014arXiv1406.0652L}.
Another approach followed to correct for this effects is  to  apply to the theoretical (model) frequencies, 
the empirical corrections obtained  by \citet{2008ApJ...683L.175K} from the seismic solar model,
\begin{eqnarray}
\label{nearsurf}
\nu_{n, l}^\mathrm{mod, corr} \, = \, \nu_{n, l}^\mathrm{mod}+
\frac{a_\mathrm{SE}}{r_\mathrm{SE}}
\left(\frac{\nu_{n, l}^\mathrm{obs}}{\nu_\mathrm{max}}\right)^{b_\mathrm{SE}} \, ,
\end{eqnarray}
where $\nu_{n, l}^\mathrm{mod, corr}$ is the corrected frequency,  $\nu_{n, l}^\mathrm{mod}$ and $\nu_{n, l}^\mathrm{obs}$ are respectively the computed and observed frequency, $b_\mathrm{SE}$ is an adjustable coefficient, $r_\mathrm{SE}$ gets close to unity when the model approaches the best solution and $a_\mathrm{SE}$ is deduced from the values of $b_\mathrm{SE}$ and $r_\mathrm{SE}$. 
This approach has been followed by, for instance, \citet{2008ApJ...683L.175K}, 
\citet{2011A&A...527A..37B}, \citet{2011a&a...535a..91d}, and \citet{2014arXiv1406.0652L}.
\citet[][]{2012ApJ...749..109G} also considered these corrections, in a Bayesian approach.

\cite{2008ApJ...683L.175K} obtained a value of $b_\mathrm{SE ,\odot}=4.9$ 
when adjusting the relation on solar radial modes frequencies.
However the value of $b_\mathrm{SE ,\odot}$ should depend on the input physics in the solar model
 considered. Indeed, \cite{2011a&a...535a..91d} obtained $b_\mathrm{SE, \odot}=4.25$ for
  a solar model computed with a different stellar evolution code and adopting a different  
  description of convection. Moreover, $b_\mathrm{SE}$ could  differ from one star to another. 
  If precise seismic observations are available to constrain
  a given star modelling, it is possible to treat $b_\mathrm{SE}$ as a variable parameter of the modelling to be adjusted 
  to minimize the differences between observed and computed individual frequencies \citep{2014arXiv1406.0652L}.

   \subsubsection{Data correlations}
   \label{correl}

    \begin{figure}
 \begin{center}
\includegraphics[width=0.9\textwidth]{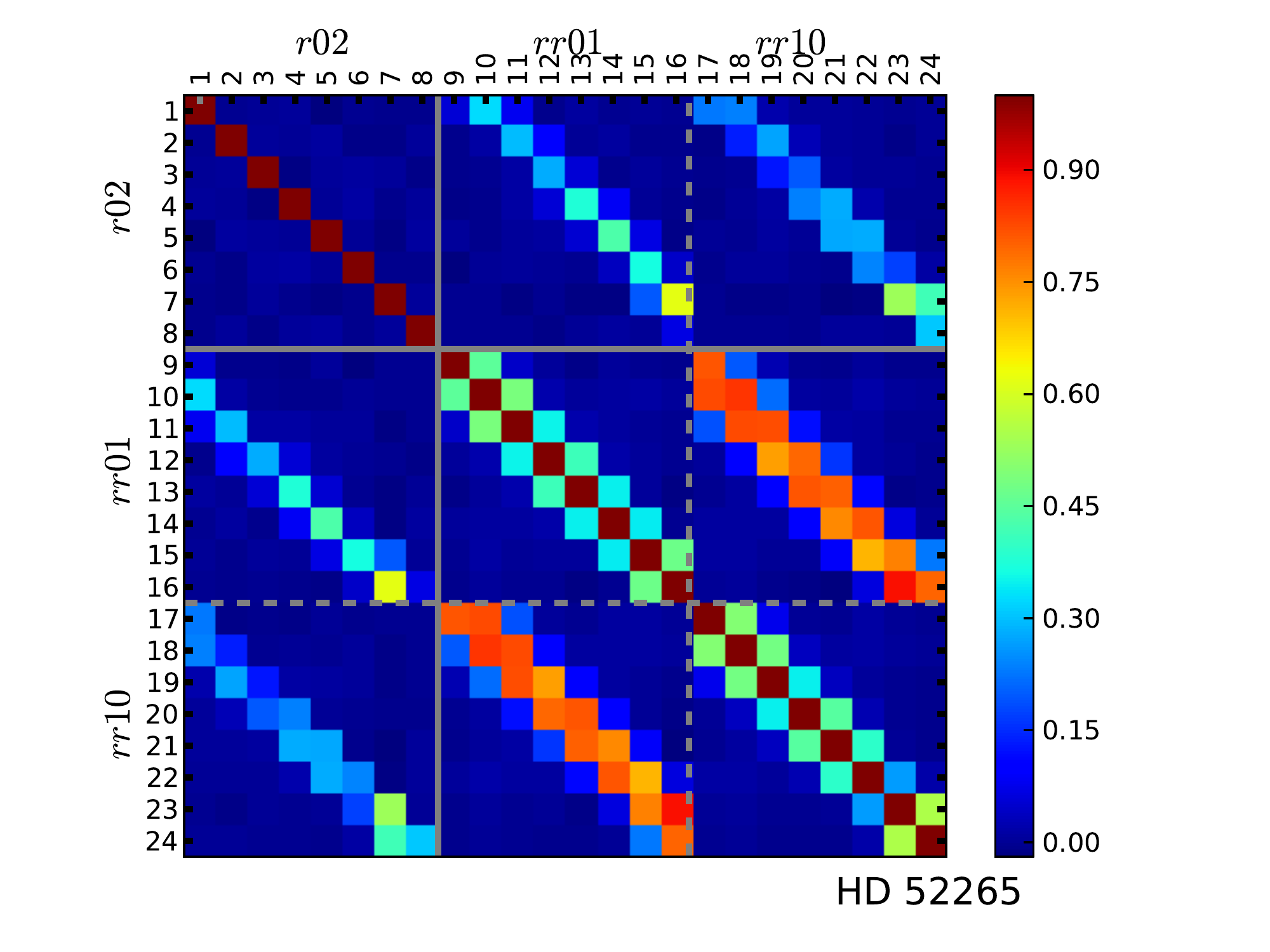}
      \caption{
 Seismic data correlations for \textsl{CoRoT} target HD~52265. Elements of the correlation matrix of the observed ratios $r_{02}(n)$ 
 and $rr_{01/10}(n)$, evaluated by a Monte Carlo simulation. Lines (columns) 1 to 8 correspond to the $r_{02}(n)$ ratios ($n$ is 
 in the range 17-24).  Lines (columns) 9 to 16 correspond to the $rr_{01}(n)$ ratios and lines (columns) 17 to 24 correspond to 
 the $rr_{10}(n)$ ratios ($n$ is in the range 16-23). As expected, there are strong correlations between some data, in 
 particular between the $rr_{01}(n)$ and $rr_{10}(n)$ ratios with the same $n$ value or consecutive values of $n$. 
 [From \citet{2014arXiv1406.0652L}.]}
 \label{correlations}
 \end{center}
 \end{figure}
    
   In the procedure of matching observational seismic data to theoretical data, one has to be careful about the possible data correlations.
   This is the case, for instance when stellar models are optimized, using not the individual oscillation frequencies, 
but some frequency differences or frequency ratios.
   In particular, the individual separations ratios $r_{02}(n)$ and $rr_{01/10}(n)$, 
are strongly correlated as illustrated in Fig.~\ref{correlations}.
 The $\chi^2$ has then to be calculated according to 
   \begin{eqnarray}
   \label{chi2cor}
   \chi^2 \; = \; \sum_{i=1}^{N_\mathrm{obs}} \left ( x_\mathrm{i, mod} - 
x_\mathrm{i, obs} \right )^\mathrm{T} \, \mathbf{C}^{-1} \,  \left ( x_\mathrm{i, mod}
 - x_\mathrm{i, obs} \right ) \, ,
   \end{eqnarray}
   where $\mathbf{C}$ denotes the correlation matrix and $T$ a transposed matrix \citep{2002nrc..book.....p}.
   
\section{Applications}
\label{applications}

\subsection{Main sequence solar-like oscillators, pressure modes}

\subsubsection{A calibrator, HD~52265}
\label{HD52265}
  
\begin{figure}[!ht]
 \begin{center}
\includegraphics[width=0.8\textwidth]{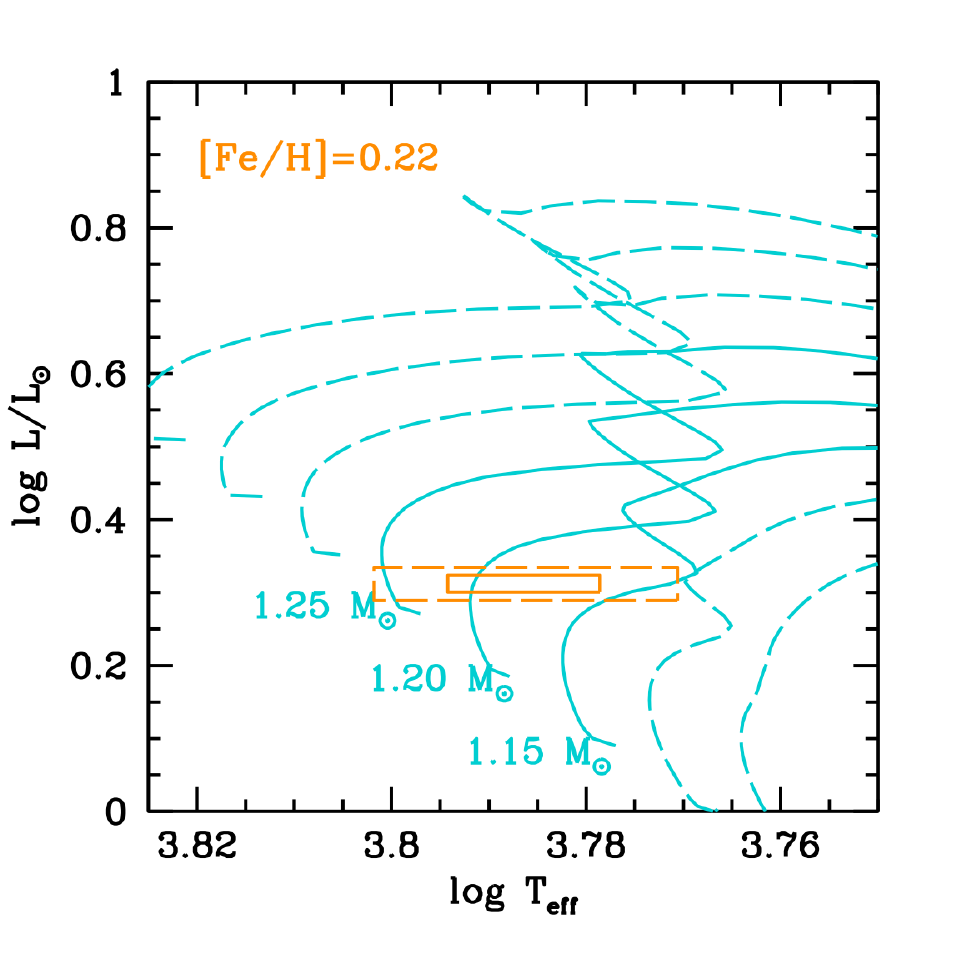}
 \caption{Position of HD~52265 in the HR diagram together with evolutionary tracks of different masses
  corresponding to its observed metallicity.
 The inner (resp. outer) error box corresponds to $1\sigma$ (resp. 
$2\sigma$) errors.}
 \label{HR_HD52}
 \end{center}
    \end{figure}

HD~52265 was one of \textsl{CoRoT} seismic main targets. It is a solar-like oscillator, metal-rich, MS star. 
It hosts an exoplanet discovered by 
\citet{2000apj...545..504b} after radial velocity measurements, but the transit of which is not observable. 
This star is quite close (${\sim}29$ pc). Its classical parameters ($L$, $T_\mathrm{eff}$, and surface [Fe/H]) have been measured with a good precision.
They place the star on the MS with a mass around 
$1.2\ M_\odot$ (see Fig.~\ref{HR_HD52}).

\citet{2011a&a...530a..97b} analysed the \textsl{CoRoT} light-curve and identified 31 pressure modes with angular
    degrees $\ell= 0-2$. The errors on the frequencies
 are small, in the range $0.2-1.2 \, \mu$Hz. 
    The mean large frequency separation is
 $\langle\Delta\nu\rangle=98.3 \pm 0.1 \, \mu$Hz 
    and the frequency at maximum power 
$\nu_\mathrm{max}= 2090 \pm 20  \, \mu$Hz.

Because of the number and good precision of the observational constraints available for this star, 
HD~52265 can be considered as a calibrator. Therefore, it can be modelled practically as a Sun, \ie, an in-depth modelling
taking into account both its classical and seismic observational constraints is possible.

An \`{a} la carte approach has been followed by \citet{2013eas63123, 2014arXiv1406.0652L} to model the star. 
The star has also been modelled by 
\citet{2012A&A...543A..96E}. \citet{2014arXiv1406.0652L} studied the different levels of 
refinement possible in the modelling of the star.
Such study allows to estimate the level of precision and accuracy on the age-dating and weighing that would be reachable for
different stars with different sets of observational constraints. 
\citeauthor{2014arXiv1406.0652L} modelled HD~52265 as a case-study, considering the following possible cases:
\begin{itemize}
\item Case $a$: a star for which only the classical parameters are measured.
\item Case $b$: a star for which the classical parameters and the mean large frequency separation $\langle \Delta \nu \rangle$ 
are measured.
\item Case $c$: a star for which the classical parameters, $\langle \Delta \nu \rangle$  and the mean small frequency separation
$\langle d_{02}\rangle$  are measured.
\item Case $d$: a star for which the classical parameters and the individual frequency separation ratios 
$r_{02}(n)$ and $rr_{01/10}(n)$ are measured.
\item Case $e$: a star for which the classical parameters and the individual frequencies $\nu_\mathrm{n, \ell}$ are available.
\end{itemize}

For each case, \citet{2014arXiv1406.0652L}, calculated a model of HD~52265 
optimized so as to match the observational constraints.
The more observational constraints, the more free parameters of the model can be adjusted in the optimization process.
We present here only a subset of these optimizations. We list in Table~\ref{Cases} the 
observational constraints considered and the model parameters that are adjusted.
Also, for each case, several models were optimized assuming different input physics.
First, a reference set of input physics was chosen (REF set, described in Sect.~2.4.3 of Lecture 1).
Then, other sets were considered changing, one at a time, the inputs of the models.
These different sets are summarized in Table~\ref{sets}, but see \citet{2014arXiv1406.0652L} for more details.

\begin{table}
\begin{center}
\caption{Sets of observational constraints considered in the optimization of the model 
of HD~52265, and the free parameters that can be adjusted accordingly, extracted from the full analysis of
\citet{2014arXiv1406.0652L}. Case $2$ is Case $2$b in \citeauthor{2014arXiv1406.0652L}. 
In Cases $a$ and $b$, $Y_0$ has to be fixed to a chosen value or derived from $(\Delta Y/\Delta Z)_\odot$ (see Sect. 3.1.2 in Lecture 1). 
In Case $b$, $\alpha_\mathrm{ conv}$ has to be fixed to a chosen value or to the solar model value (Sect. 3.1.2, Lecture 1).
The age and mass are denoted by $A$ and $M$.}
\label{Cases}
\begin{tabular}{llll}
\hline\hline
Case & Observed  &Adjusted  & Fixed \\
\hline
$a$                     &  $T_\mathrm{{eff}}$, $L$, [Fe/H]                                                 & A, M, $(Z/X)_0$ & $\alpha_\mathrm{ conv}$, $Y_0$\\
$b$ &  $T_\mathrm{{eff}}$, $L$, [Fe/H], $\langle\Delta\nu\rangle$ & A, M, $(Z/X)_0$, $\alpha_\mathrm{ conv}$ & $Y_0$\\
$c$      &  $T_\mathrm{{eff}}$, $L$, [Fe/H], $\langle\Delta\nu\rangle$, $\langle d_{02}\rangle$  & A, M, $(Z/X)_0$, $\alpha_\mathrm{ conv}$, $Y_0$ & --\\
$d$                    &  $T_\mathrm{{eff}}$, $L$, [Fe/H], $r_{02}(n)$, $rr_{01/10}(n)$ & A, M, $(Z/X)_0$, $\alpha_\mathrm{ conv}$, $Y_0$ & --\\
$e$                     &  $T_\mathrm{{eff}}$, $L$, [Fe/H], $\nu_{n, \ell}$                     & A, M, $(Z/X)_0$, $\alpha_\mathrm{ conv}$, $Y_0$ & --\\
\hline
\end{tabular}
\end{center}
\end{table}
\begin{table}  
\caption{Sets of input physics considered by \citet{2014arXiv1406.0652L} 
in the modelling of HD~52265. As detailed in Lecture 1, 
the reference set of inputs denoted by REF is based on \textsl{OPAL05} EoS, 
\textsl{OPAL96/WICHITA} opacities, \textsl{NACRE+LUNA}
  reaction rates (this latter only for $^{14}N(p,\gamma)^{15}O$), the 
CGM formalism for convection, the MP93 formalism for microscopic diffusion, 
the Eddington grey atmosphere, and \textsl{GN93}
 solar mixture. It includes neither overshooting, nor convective penetration
 or rotation. For the other cases we only indicate the input that is changed
 with respect to the reference. 
Colours and symbols in column 3  are used in Fig. \ref{age}.}
\vspace*{0.4cm}
\begin{center}
\begin{tabular}{lll}
\hline\hline
Set & Input physics & Figure symbol/colour \\
\hline
$A$ & REF  &  circle , cyan \\
$B$ & convection MLT  & square,  orange\\
$C$ & AGSS09 mixture  &  diamond, blue\\
$D$ & NACRE for $^{14}N(p,\gamma)^{15}O$ & small diamond, magenta\\
$E$ & no microscopic diffusion  & pentagon, red\\
$F$  &Kurucz model atmosphere, MLT  & bowtie, brown\\%
$G$  & B69 for microscopic diffusion  & upwards triangle, chartreuse\\
$H$  & EoS OPAL01  & downwards triangle, purple\\
$I$   & overshooting $\alpha_\mathrm{ ov}{=}0.15 H_P$  & inferior, yellow\\%
$J$   & overshooting $M_\mathrm{ ov, c}{=}1.8\times{M_\mathrm{ cc}}$   & superior, gold  \\%
$K$  & penetrative convection $\xi_\mathrm{PC}{=}1.3 H_P$  & asterisk, pink\\
\hline
\end{tabular}
\end{center}
\label{sets}
\end{table}

\begin{figure}[!ht]
 \begin{center}
\includegraphics[width=0.9\textwidth]{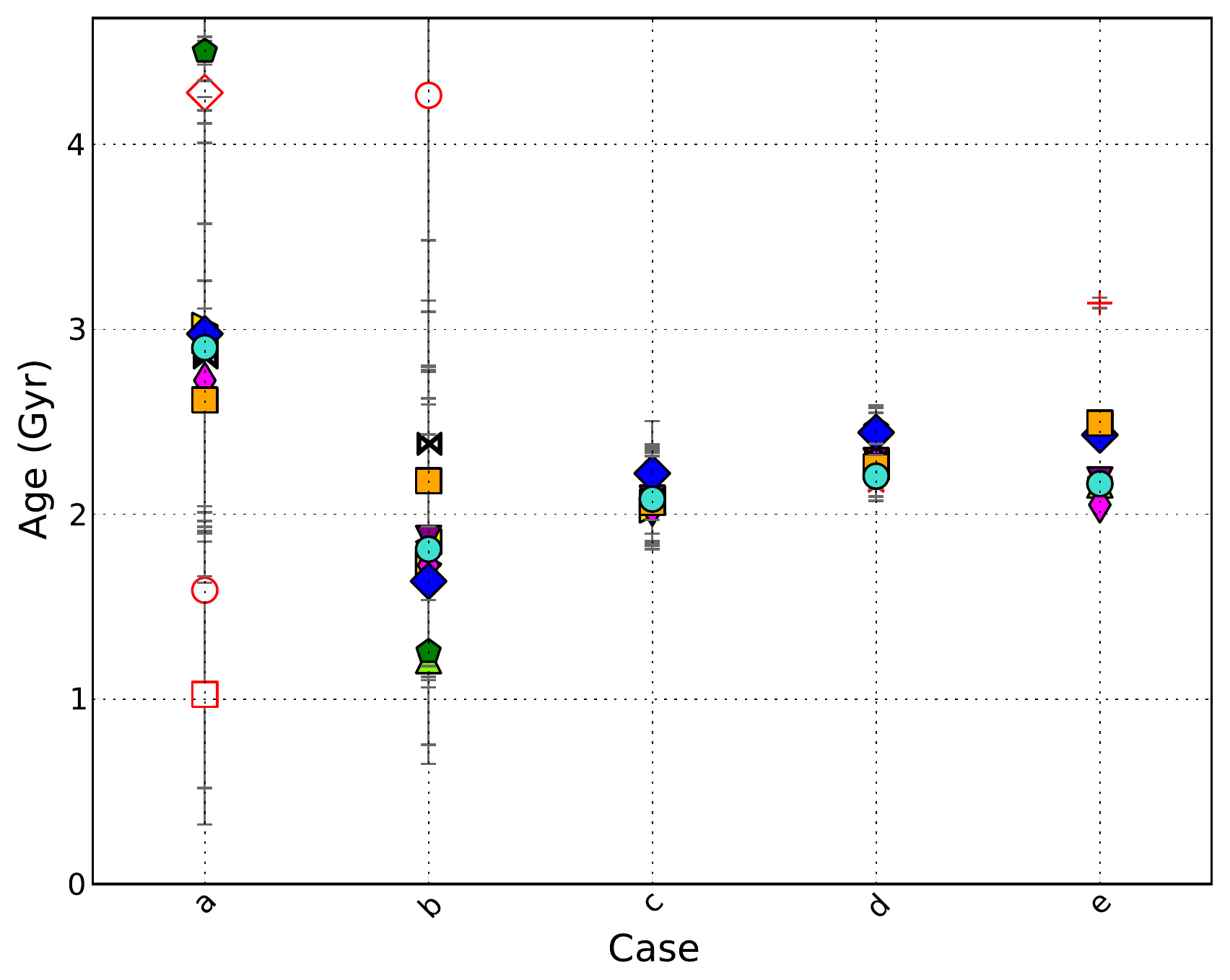}
 \caption{The ranges of ages derived from stellar model optimization for HD~52265. In abscissae
  are listed the cases numbers, as defined in Table~\ref{Cases}. For each case, 
  several model optimizations can be identified according to the symbols and colours indicated in Table~\ref{sets}. 
  In addition, open red symbols are for extra models. 
  In cases $a$ and $b$, 
 circles correspond to $Y_0$ values smaller than the one inferred from $(\Delta Y/\Delta Z)_\odot$. 
 In case $a$, square and diamond are for $\alpha_\mathrm{conv}$ values changed by 
  ${\pm}0.20$ dex with respect to solar.
 In case $e$, the red cross is a model without corrections from surface 
effects.}
 \label{age}
 \end{center}
    \end{figure}

In Fig.~\ref{age}, we show the age range corresponding to the optimization cases in Table~\ref{Cases}.
Case $a$ shows a large scatter in age, which results from the large range of possible values of the free parameters, 
\ie, 
 initial helium content $Y_0$ and mixing-length convection parameter $\alpha_\mathrm{conv}$.
In Case $b$, $\alpha_\mathrm{conv}$ is inferred from the model optimization, but
$Y_0$ still is a free parameter, so the age scatter remains large. Cases $c$, $d$, and 
$e$ are constrained by seismology.
These cases show a spectacular improvement of the precision and accuracy on the age-dating of HD~52265, due to the fact
that when seismic constraints are available, the initial helium content and mixing-length parameter are 
constrained together with the age, mass, and initial $Z/X$.  Table~\ref{recap} provides a summary 
of the precision on the age in the different cases.
We point out that, although cases $d$ and $e$ have similar ages (and age error bar), 
case $d$ should be preferred because no surface effects corrections
of the frequencies are needed when considering the frequency separation ratios as constraints. Indeed, in case $e$, 
if the individual frequencies
are not corrected, the age is larger by $40$ per cent (red cross in Fig.~\ref{age}). 

Also, the optimizations show that there is a range of possible  values of the ($Y$, $M$) pair. 
This mass-helium degeneracy does not impact the age-dating but hampers the precise
determination of the mass. Concerning the determination of the mass and radius 
of HD~52265, \citet{2014arXiv1406.0652L} conclude that seismology allows to reach a 
precision of $7$ per cent on mass and
$3$ per cent on radius.

 \begin{table}
 \begin{center}
 \caption{Age of HD~52265 obtained with different sets of observational constraints.}
 \label{recap}
 \vspace*{0.4cm}
 \begin{tabular}{ccc}
 \hline\hline
 Case & $A$  (\Gyr) & $\Delta A/A$ (\%) \\
 \hline
$a$ & $3.42\pm3.10$  & $90$  \\
$b$ & $3.28\pm2.52$ & $77$   \\
$c$ & $2.15\pm0.35$ &$16$  \\
$d$ & $2.31\pm 0.24$ &$10$   \\
$e$ & $2.27\pm 0.24$ & $10$   \\
 \hline
 \end{tabular}
 \end{center}
 \end{table}
 
 Noteworthy a by-product of  \`{a} la carte stellar modelling based on asteroseismic constraints is the
 determination of a seismic stellar gravity, $\log g_\mathrm{seism}$. 
 The seismic gravity has been shown to be quite insensitive to the model input physics and to be 
 much more precise than the spectroscopic one \citep[\eg,][]{2012ApJ...749..152M,2012ApJ...748L..10M,2012MNRAS.419L..34M}.
 For instance, for HD~52265, the error bar on the spectroscopic gravity 
is of ${\sim}0.2$ dex, 
 while the one on $\log g_\mathrm{seism}$ is $0.02$ dex, \ie, ten times 
smaller. 
 Therefore, once the seismic $\log g_\mathrm{seism}$ is obtained via a 
stellar model optimization,
 it is possible to use it in a reanalysis of the star spectrum, and, in 
turn to improve the determination of the
 effective temperature and metallicity of the star.
This technique has been applied to the spectroscopic analysis of 
 \textsl{CoRoT} targets by \citet{2013A&A...552A..42M} (two MS stars) and
 by \citet{2014A&A...564A.119M} (19 RGs).
Moreover, $\log g_\mathrm{seism}$ is adopted as a calibrator in pipelines
 of the large spectroscopic surveys \textsl{APOGEE} and 
Gaia-ESO-Survey. It was also proposed for the $\log g$ 
determination of \textsl{Gaia} stars \citep{2013MNRAS.431.2419C}.
Such a back and forth analysis, 
 combining stellar interior and atmosphere models, is expected
 to provide much better models of stars.
 
\subsubsection{An exoplanet host with observed planetary transit, HD~17156}
\label{HD17156}

 \begin{figure}[!hp]
 \begin{center}
\includegraphics[width=0.7\textwidth]{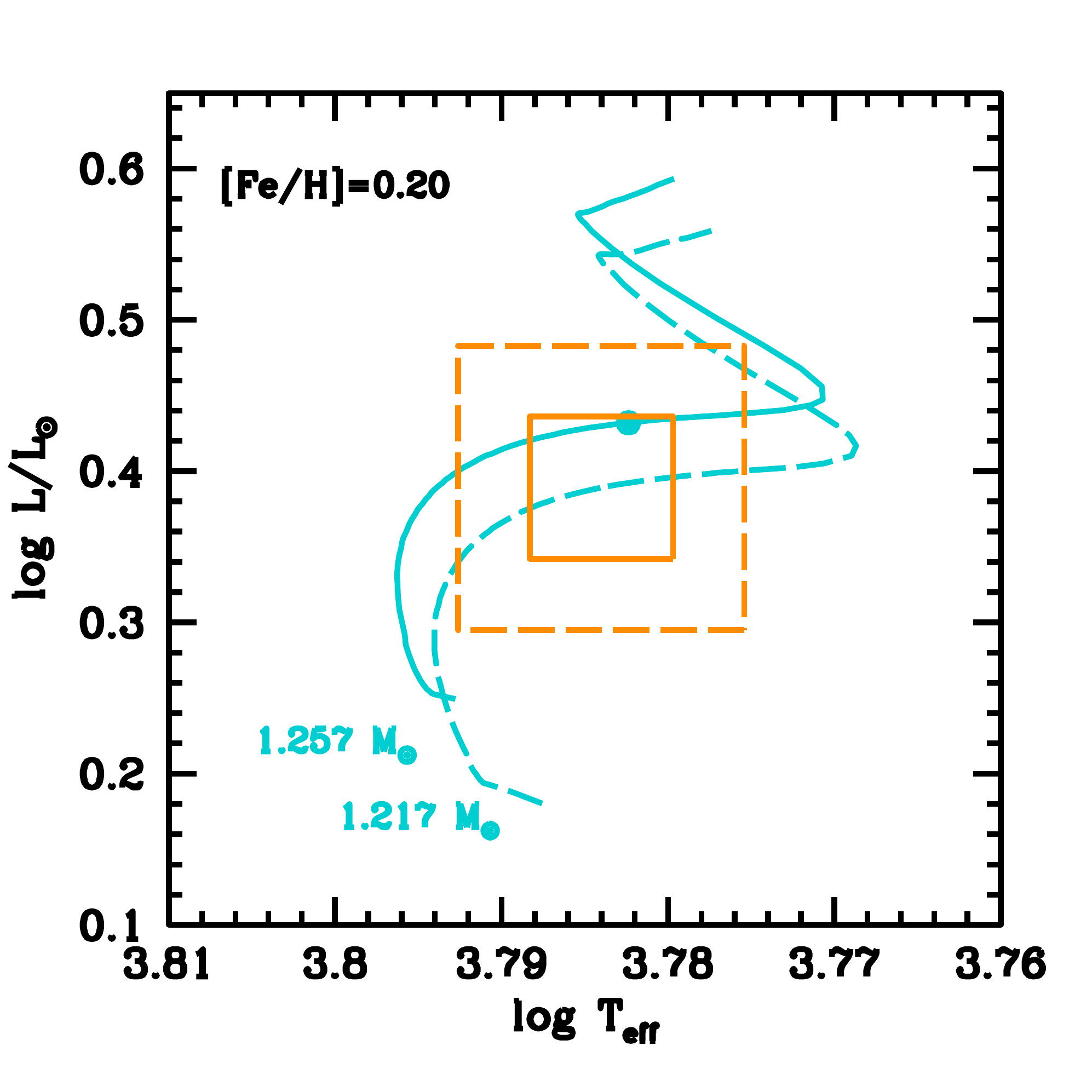}
\includegraphics[width=0.7\textwidth]{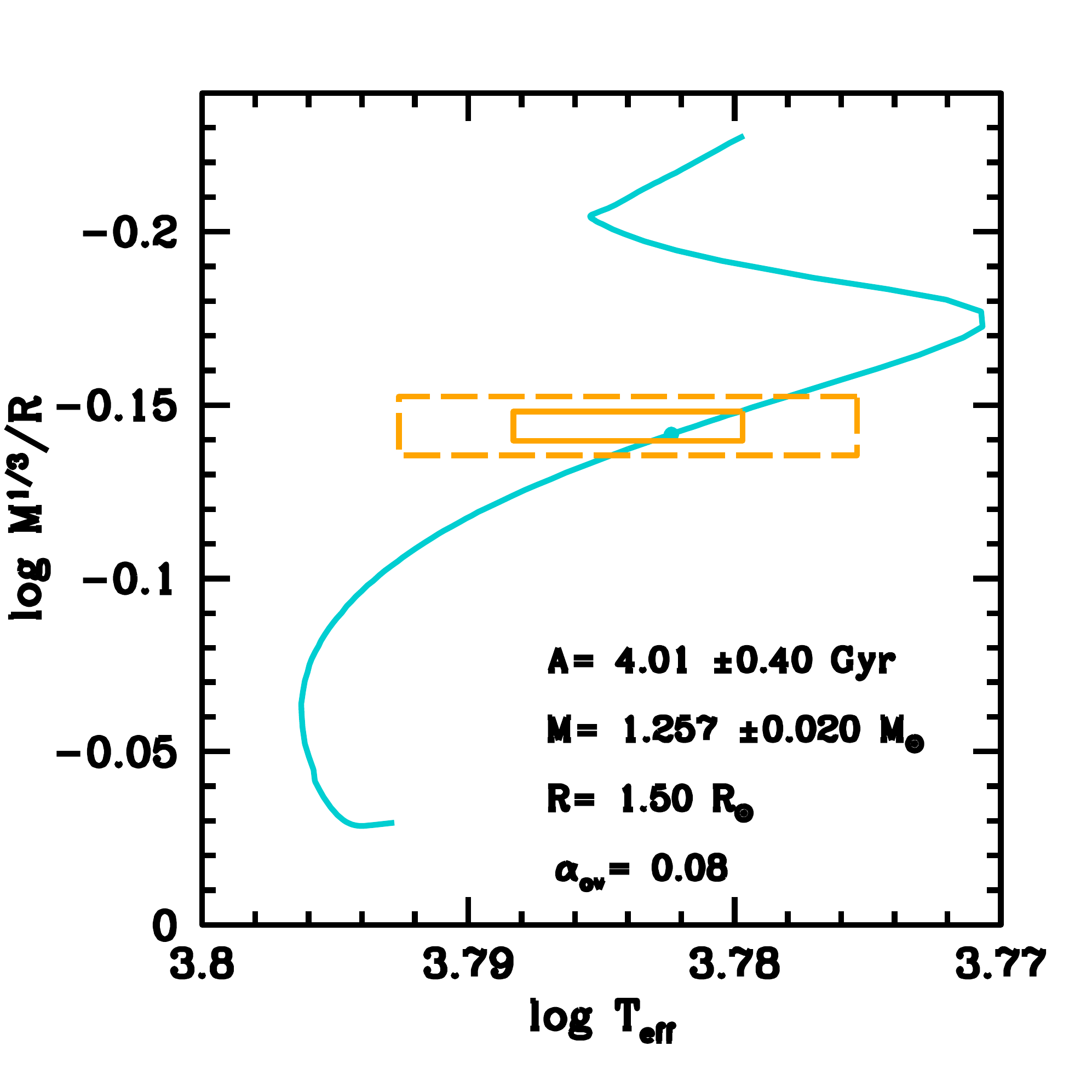}
 \caption{The exoplanet host, HD~17156. 
 {\sl Top}: HR diagram. {\sl Bottom}: $M^\frac{1}{3} R^{-1} - T_\mathrm{eff}$ 
diagram.  
 The dashed-line is a model optimized with the classical parameters only, 
the
 continuous line  is a model optimized with classical, seismic, and transit 
constraints.
 The inner (resp. outer) error box corresponds to $1\sigma$ 
(resp. $2\sigma$) errors.
[From \citet{2012ASPC..462..469L}.]}
 \label{HD17156HR}
 \end{center}
    \end{figure}
 
The characterisation of exoplanet host stars, and in turn of their exoplanet, can be improved if, 
in addition to seismic constraints, the transit of the exoplanet is observed. Indeed, the measure of the
duration of the different phases of the transit together with the third Kepler law, 
provide the quantity $M_\star^\frac{1}{3} R_\star^{-1}$ itself proportional to
$\langle \rho \rangle^{1/3}$ ($\langle \rho \rangle$ is the mean density). 
Since the mean large frequency 
separation is related to $\langle \rho \rangle^{1/2}$, 
two independent measurements of the mean stellar density are available in that case.

One example is the star HD~17156, observed by the HST and modelled by  \citet{2011apj...726....3n}, \citet{2011ApJ...726....2G} and \citet{2012ASPC..462..469L}. 
Like HD~52265, it is a metal rich star on the MS, in the same domain of mass. 
The quantity $M_\star^\frac{1}{3} R_\star^{-1}$ was obtained from the transit together with
eight oscillations frequencies (less precise and numerous than those of HD~52265).
In Fig.~\ref{HD17156HR} we illustrate how seismic and transit measurements 
change the optimized model of HD~17156. The left panel shows the HR diagram  and the 
right panel shows the $M^\frac{1}{3} R^{-1} - T_\mathrm{eff}$ plane. An optimization based on the classical parameters
only, assuming solar $\alpha_\mathrm{conv}$ and $(\Delta Y/\Delta Z)_\odot$,
matches the HR position and metallicity for a mass of $M= 1.22 \pm 0.02 \, M_\odot$ and an age of $A=4.33\pm0.40$~\Gyr.
On the other hand, if seismic and transit constraints are also considered 
the best model has 
$M= 1.26 \pm 0.02 \, M_\odot$ and an age of 
$A=4.01\pm 0.40$~\Gyr~\citep{2012ASPC..462..469L}.

\subsection{Advanced stages, mixed modes}
\label{mixed}

\subsubsection{A subgiant star, HD~49385}

As described in Sec.~\ref{pgmode},  the increase of density in the
central regions due to the core contraction at the end of  the MS phase and subsequent evolution in the subgiant branch leads to an
 increase of $N_\mathrm{BV}$ in these regions  and therefore of the frequencies of $g$-modes.  The range of frequencies of $g$-modes may 
 overlap that of $p$-modes, and if the evanescent region is narrow enough $p$- and $g-$modes of the same degree  can interact. This 
 interaction results in modes with a mixed $g-p$ nature that propagate in the inner and outer layers and whose frequency  may 
 significantly
deviate from the regular spacing between consecutive overtones characteristic of pure $p$-modes.  The presence and identification of this kind of modes provides a  stringent constraint on the internal properties of the star and on its evolutionary state.  Although its potential has been already investigated in the case of
$\eta$ Bootis \citep{2004SoPh..220..185D},   12 {Bootis}
\citep{2007MNRAS.377..373M}, and $\beta$~Hydri \citep{2011A&A...527A..37B}, an unambiguous identification of mixed
modes and its use in fitting stellar parameters became  possible
only recently with the high precision photometric data provided
by \textsl{CoRoT} and  \textsl{Kepler} \citep{2010A&A...515A..87D,2011a&a...535a..91d,2010ApJ...723.1583M,2011a&a...535a..91d,2011ApJ...733...95M,2012ApJ...745L..33B, 2013ApJ...763...49D}.

\begin{figure}[!htp]
 \begin{center}
\includegraphics[width=0.7\textwidth]{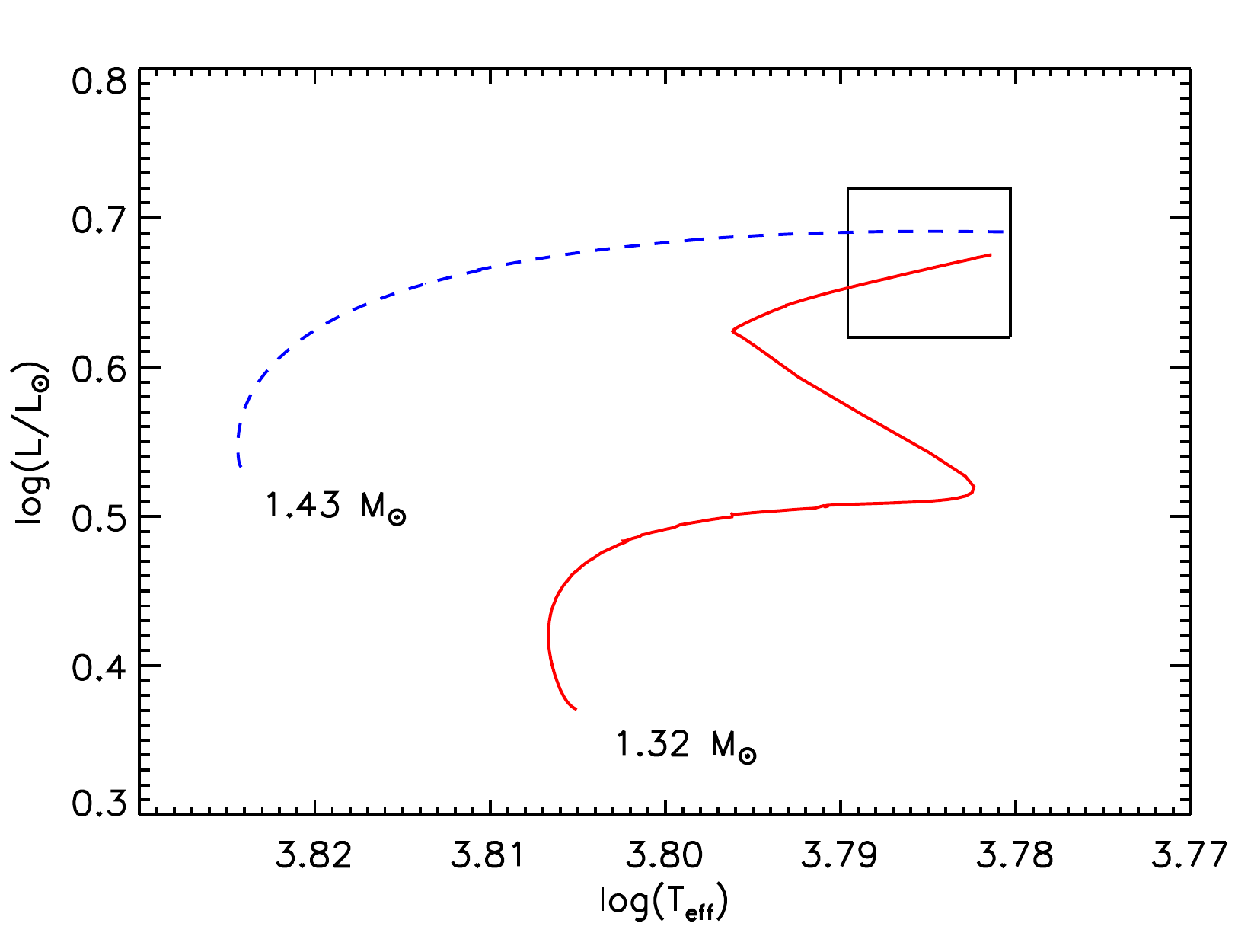} 
\includegraphics[width=0.7\textwidth]{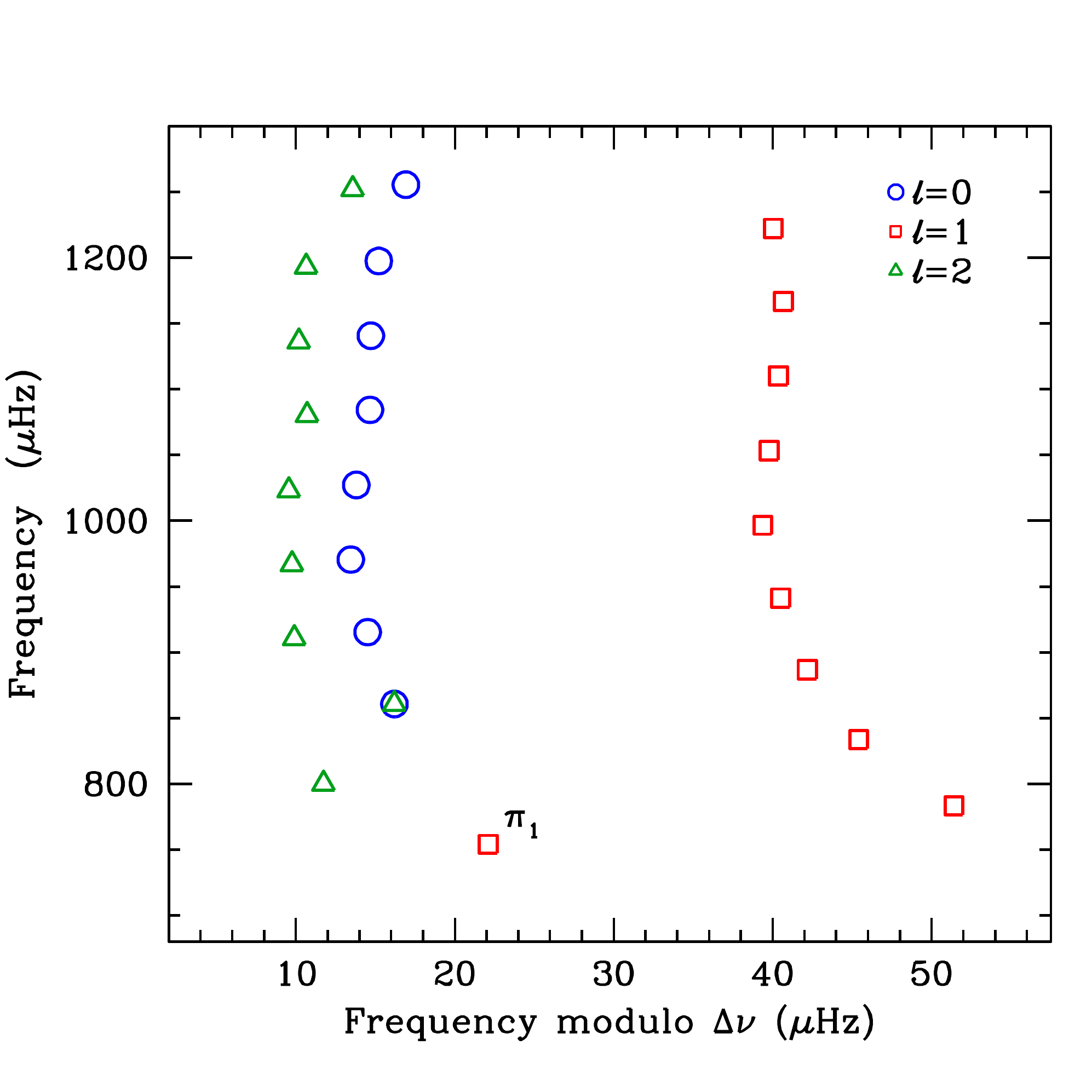}
 \caption{{\sl Top:} {evolutionary tracks of two models fitting the
   position of HD 49385 in the HR diagram. The box corresponds to the
   temperature and luminosity observed values with 1-$\sigma$ error
   bar. The dashed line corresponds to a MS model and the solid line to a
   post-MS one.} {\sl Bottom:} \'echelle diagram of HD~49385 for 
   radial (blue circles), dipolar (red squares),  and quadruple  (green
   triangles) modes fitted in \citet{2010A&A...515A..87D}. The mode labelled $\pi_1$ is
   undergoing an avoided crossing.  
[From \citet{2011a&a...535a..91d}.]}
   \label{echell_hd49385}
   \end{center}
 \end{figure}

HD 49385 is a G0-type star with an apparent magnitude of $m_{\rm V}=7.39$ \citep{1998A&AS..129..431H} 
which was observed by \textsl{CoRoT} over a period of 137
 days. Its atmospheric  parameters ($T_{\rm eff}=6095\pm65 \, K$, $\log g=4.0\pm0.06$, and
[Fe/H]=$0.09\pm0.05$~dex) were derived from high-quality spectra obtained with the
NARVAL spectrograph, and its {Hipparcos} parallax (13.91~mas) is
known with a precision of 5\%. The seismic and spectroscopic
characterization of this target were presented in
\citet{2010A&A...515A..87D}.

The seismic analysis of the \textsl{CoRoT} light curves showed solar-like
oscillations with a frequency at the maximum  power $\nu_{\rm
  max}=1010 \pm 10\,\mu$Hz, and a mean large frequency separation of
radial modes of $\langle\Delta\nu\rangle=56.3\pm0.5\,\mu$Hz. In the
domain around 1~mHz a series of peaks associated to $p$-modes of degree
$\ell=0-2$ over nine radial orders were identified 
(Fig.~\ref{echell_hd49385}, top panel). 
Moreover, thanks to the high quality of the spectrum it was also possible to detect a peak outside the 
ridges that could be associated to a $\ell=1$ mixed mode ($\pi_1$). 
Although three very clear ridges appear in the \'echelle diagram of the power spectrum, 
some of the modes do not follow the expected pattern of high-order-radial p modes, 
and the curvature of the $\ell=1$ ridge at low frequency significantly  differs from that of the $\ell=0$ ridge. 
These features are the signatures of the presence of a low-degree mixed mode undergoing an avoided crossing \citep[\eg,][]{2010Ap&SS.328..259D}. 

The presence of these seismic features also allows for the
classification of HD 49385 as a subgiant. In fact, mixed modes appear
in the solar-like oscillation domain only for models evolved enough to
bring  the frequency of the $g$-mode undergoing the avoided crossing to
the $p$-mode frequency domain. Although the low value of $\log g$ from
spectroscopy already suggests that HD 49385 could be an evolved star,
the degeneracy between MS and Post-MS models  hinder the
identification of their evolutionary state solely based  on the values
of classical parameters. {Figure~\ref{echell_hd49385} (top panel)  shows the evolutionary
  tracks of two models fitting the position of HD 49385 in  the HR
  diagram  and the large frequency separation for radial
  modes. Although the MS model can reasonably well fit the $\ell=0$
  ridge, it is not able to match the curvature of the $\ell=1$ ridge
  shown in the right panel.}

The frequency of the $g$-modes undergoing the avoided crossing, and the
curvature of the $\ell=1$ ridge contain information directly
linked to the age and properties of the inner regions of the
star. The deviation of the frequencies with respect to the expected
ridges depends on the strength of the coupling between acoustic and
gravity mode cavities, and then, on the value of $N_{\rm BV}$ in the
evanescent zone.

The rapid variation of the frequency of modes affected by avoided
crossing converts the detection of these modes into  a 
powerful tool to constrain stellar age. However, that rapid
variation also implies that the usual approach of star modelling
consisting in finding the optimal model that minimizes a merit
function $\chi^2$  for the  model frequencies may become tremendous
time-consuming. To overcome this difficulty \citet{2011a&a...535a..91d}
suggested to adapt the traditional grid-of-model approach and use  as main seismic constraints  the frequency of
the mode undergoing the avoided crossing ($\nu_{\rm cross}$) and the mean large frequency separation of radial modes ($\langle\Delta\nu\rangle$).

The frequency at which the avoided crossing occurs corresponds to the
frequency of the pure $g$-mode, and therefore its value is determined by
the profile of $N_{BV}$ in the $g$-cavity. 
After the exhaustion of H in
the core, at the end of MS, the inner region of the star consists in an
inert He-core  with a radiative stratification, and hence the
evolution of $N_{\rm BV}$ there is mainly determined by $\rho_{\rm
  c}$. The inner part of the star continues
contracting until the quiet central He-burning starts, and so, the
avoided crossing frequencies monotonically increase during the
evolution of the star.  On the other hand,   $\langle\Delta\nu\rangle$ monotonically decreases with age since stellar radius increases.  
As a consequence, for a given physics, $\langle\Delta\nu\rangle$ and
$\nu_{\rm cross}$  alone, without the contribution of any other classical or seismic constraints, provide an estimate of the stellar mass and age with a very high internal precision.

In practice, the value of $\nu_{\rm cross}$  is not known, and for fitting models, the frequency of the model with the highest g behavior is used.
To characterize HD 49385, \citet{2011a&a...535a..91d} computed two grids of models, one assuming \textsl{GN93} solar mixture, and the other the \textsl{AGS05} one. Moreover, for each grid, they varied the model input free parameters:
\begin{itemize}
\item Initial helium mass fraction ($Y_0$): between 0.4 and 0.28 with $\Delta Y_0=0.01$.
\item Convection mixing-length parameter ($\alpha_{\rm CGM}$): between 0.48 and 0.72, with $\Delta\alpha_{\rm CGM}=0.04$.
\item Heavy metal content $[Z/X]$: between 0.04 and 0.14, with $\Delta [Z/X]=0.05$.

\item Core overshooting ($\alpha_{\rm ov}$): between 0 and 0.2, with $\Delta \alpha_{\rm ov}=0.025$.
\end{itemize}

For each grid they derived  the models with mass and age satisfying 
 $\langle\Delta\nu\rangle^{\rm mod}= \langle\Delta\nu\rangle ^{\rm
   obs}=56.3\,\mu$Hz and $\nu_{\rm cross}$=
 $\nu_{\pi,\ell=1}=748.6\,\mu$Hz.   The merit function  based on individual corrected
 frequencies was calculated only for these models. 
 The behavior of $\chi^2$ as a function of $\alpha_{\rm
   ov}$ showed two minima, one at $\alpha_{\rm ov}<0.05$ and the other at
 $\alpha_{\rm ov}=0.19\pm 0.01$. The four families of solutions,
 depending on the core overshooting during MS and solar mixture,
 provided best-fit models with very similar values of the observables
 and it was not possible to discriminate between them.  In spite of
 that,  the stellar parameters are well constrained:
 $M=1.25\pm0.05\,M_\odot$  (uncertainty of 4\%);
 $R=1.94\pm0.03\,R_\odot$  (uncertainty of 1.5\%); $\log g=3.954\pm 0.009$, in good agreement 
 with spectroscopic value; and age $\tau=5.02\pm0.26$~\Gyr~(uncertainty of 5\%). 
 The  characterization of other \textsl{Kepler} subgiants by fitting the 
frequencies of mixed modes also leads 
 to uncertainties in the age of the order of 5--7\%, a significant improvement compared to the 35-50\% 
 obtained using scaling relations and grid-based modeling \citep{2013ApJ...763...49D}.

The detailed analysis of the fitting procedure used in
\citet{2011a&a...535a..91d} showed that the obtained results of the
fit seem completely determined by the curvature of the $\ell=1$
ridge. Therefore the distortion of the $\ell=1$ ridge caused by the
avoided crossing  plays a crucial role in constraining the interior of
HD 49385. This curvature is also a good estimate of the  coupling
between acoustic and gravity cavities (hence of the stellar structure
in the evanescent region) if the frequency of the avoided crossing in
the models matches that of the observations. Another interesting
result of \citet{2011a&a...535a..91d}'s study is that the parameter
describing the curvature of the $\ell=1$ ridge is strongly
anti-correlated to the stellar mass of the model \citep[see
also][]{2012ApJ...745L..33B},  reason why the mass of HD 49385 is so
tightly constrained.

\citet{2011a&a...535a..91d} also considered the effect of including microscopic diffusion 
in their grids of models. The derived age decreased by $\sim$1~\Gyr~(20\%), 
but the free parameters of the fitting did no change. 
In fact, the effect of transport processes that could modify $\nabla_\mu$ 
during the MS in the region currently occupied by the evanescent region, 
is rapidly erased as soon as the H-burning shell advances in the star. 
For that reason, only the parameters of models close to the TAMS will be potentially 
affected by the inclusion of diffusive or turbulent chemical transport. 

\subsubsection{The evolutionary stage of red giants}
\label{redgiants}

Red giants play a crucial role in stellar and galactic astrophysics since they serve as distance and age indicators.
However, except for RGs belonging to stellar clusters, their
characterisation is affected by large uncertainties. Ages and masses
are difficult to derive because  a narrow range of colours (or $T_\mathrm{eff}$) in the HR diagram 
corresponds to a wide domain of masses, chemical compositions, and evolutionary
states. Fortunately, RGs have an extended convective envelope,
and as in solar-like stars, turbulence may
stochastically excite acoustic oscillation modes. Red giants have been known to be pulsating stars for a long time:
classic Cepheids,  RR~Lyrae,  and Mira variables are examples of
high-amplitude red-giant  pulsators. However, it is only  in the early 2000s that low-mass RGs have been 
suspected to undergo solar-like oscillations. 
Although stochastic oscillations were detected in a few RGs
from ground- and space-based observations
\citep[\eg,][]{2002A&A...394L...5F, 2006A&A...448..689D, 2007A&A...468.1033B},
we had to wait for 
the high-precision photometry observations of 
 \textsl{CoRoT} and \textsl{Kepler} to confirm the detection of radial
 and
non-radial oscillations in many $G-K$ RGs in the
field and in three open clusters  \citep{2009Natur.459..398D,
  2009A&A...506..465H, 2010ApJ...713L.176B, 2010ApJ...713L.182S,
  2011ApJ...739...13S}.  Their spectra show similarities with 
MS solar-like oscillators spectra:  oscillations  appear in the spectrum as a gaussian-shaped power excess centred 
at $\nu_{\rm max}$ and showing a regular pattern characterized by the large
frequency separation $\langle\Delta\nu\rangle$. However, because
the structure of RGs is very different from that of the Sun
and solar-like stars, differences also appear in their oscillation spectra. 

As in MS solar-like pulsators, $\nu_{\rm max}$ and $\langle
\Delta\nu\rangle$ are linked to the classical stellar parameters, via scaling relations (see
Sec.~\ref{scaling}).  It has then been
 possible to derive, provided an estimate of $T_{\rm eff}$ was
available, the mass and radius of thousands of red giants
\citep{2009A&A...503L..21M, 2010a&a...517a..22m, 2010A&A...509A..77K, 2011MNRAS.414.2594H}. Such
a large number of model-independent stellar parameters for single
stars has no precedent and turns out to be a major contribution for
the studies of stellar populations, and formation and evolution of the
Galaxy
\citep{2009A&A...503L..21M, 2012rgps.book...11M, 2013MNRAS.429..423M}.
An estimation by \citet{2013ARA&A..51..353C} indicates 
that  the \textsl{CoRoT} 
and  \textsl{Kepler} giants cover a mass range from $\sim$ 0.9 to $\sim 3M_\odot$ corresponding 
to an age in the  range $\sim$ 0.3 to  $\sim  12$~\Gyr, that is spanning 
the entire Galactic history. 

\begin{figure}[!ht]
\begin{center}
\includegraphics[width=0.9\textwidth]{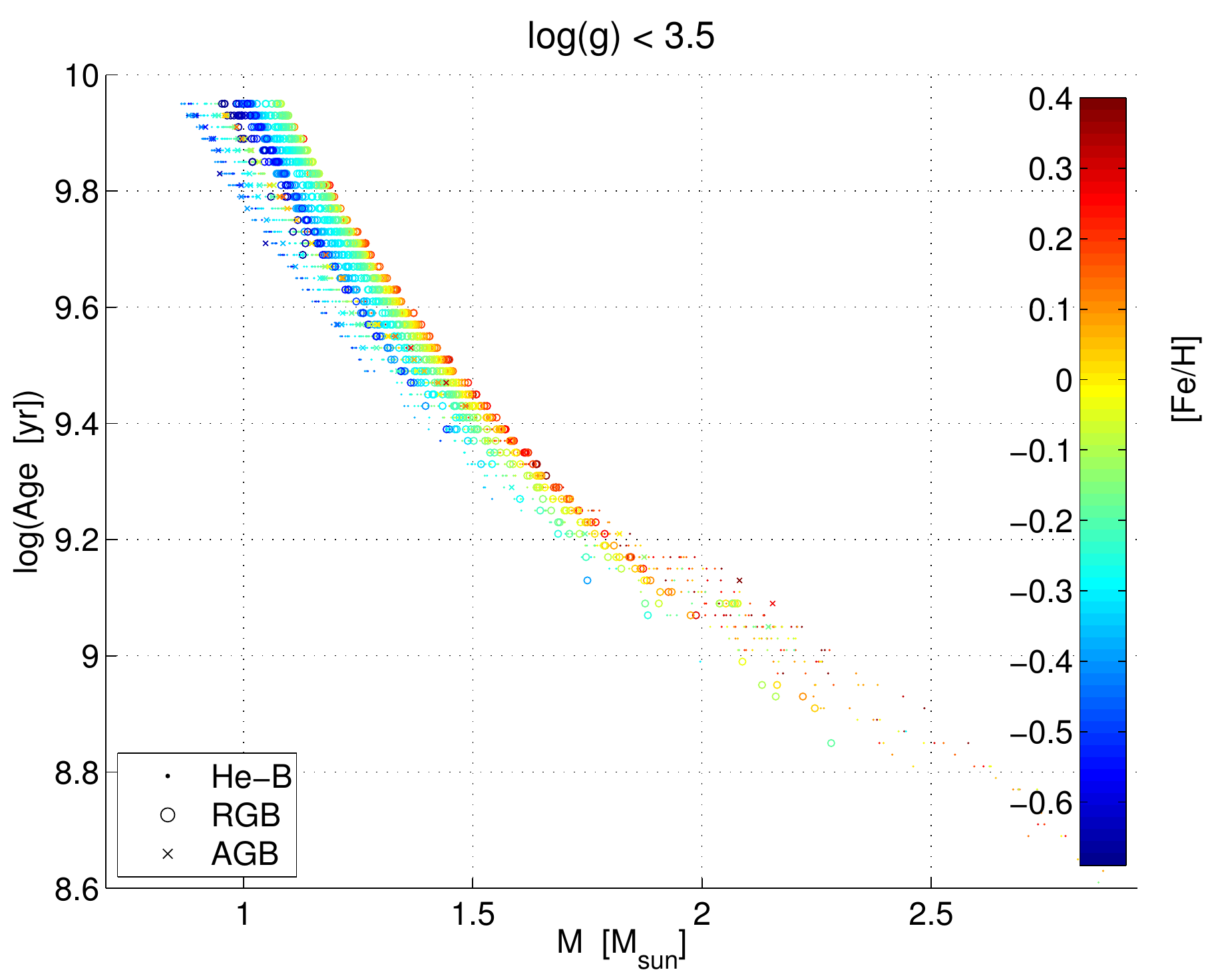}
\caption{Age-mass-metallicity relation for  red giants 
 in a synthetic population representative of thin-disc stars observed
 by \textsl{CoRoT} in the LRc01 field (first run of observation in the galactic
 centre direction). The evolutionary state of giants is marked with a different symbol: dots (stars in the 
 core-helium-burning phase, He-B), crosses (asymptotic-giant-branch stars, AGB), and open circles (stars on the
RGB). [From \citet{2012rgps.book...11M}.] 
}
\label{mass-age}
\end{center}
\end{figure}

A serious obstacle to discriminate between scenarios of formation and
evolution of the Galaxy is the difficulty of measuring distances and
ages for individual field stars. Although slightly model-dependent,
the age of a red giant is essentially determined by the time it spends in
the MS phase (and hence by its mass). Contrary to MS stars, low mass
RGs show a very tight age-mass relation (see
Fig.~\ref{mass-age}). The reason is that, while a central extra-mixing
during the central H-burning phase increases the MS lifetime (see Lecture 1), 
it also leads to a larger inert He core at the end
of the MS. This isothermal core is closer to the
Sch\"onber$g$-Chandrasekhar limit and then, the  crossing of the HR
diagram proceeds more quickly  balancing out the longer MS with a
shorter subgiant phase. So, while a 1.4~$M_\odot$ stellar model with a core overshooting
of $\alpha_{\rm ov}=0.2$ is 20\% older at TO than a model
with  no extra-mixing, it reaches the RGB at an
age only 4\% larger than that of  its non-mixed counterpart. 

The age of low-mass red giants is then very weakly affected by an
extra core-mixing during the MS and hence, once the seismic mass is derived, 
we get an estimation of the age. Figure~\ref{mass-age} shows  the
scatter in age, at a given mass, due to metallicity and evolutionary
state (shell H-burning -- RGB --, central He-burning -- He-B --, or
shell He-burning -- AGB).  Metallicity can be derived from
spectroscopy and/or photometry, but,  how can we discriminate
between different evolutionary states for  RGs in the field?  
The detailed properties of the oscillation modes depend on the stellar structure, and in turn the 
information required to discriminate between evolutionary states of RGs is contained in their oscillation spectrum. 

\begin{figure}[!hpt]
 \begin{center}
\includegraphics[width=0.7\textwidth]{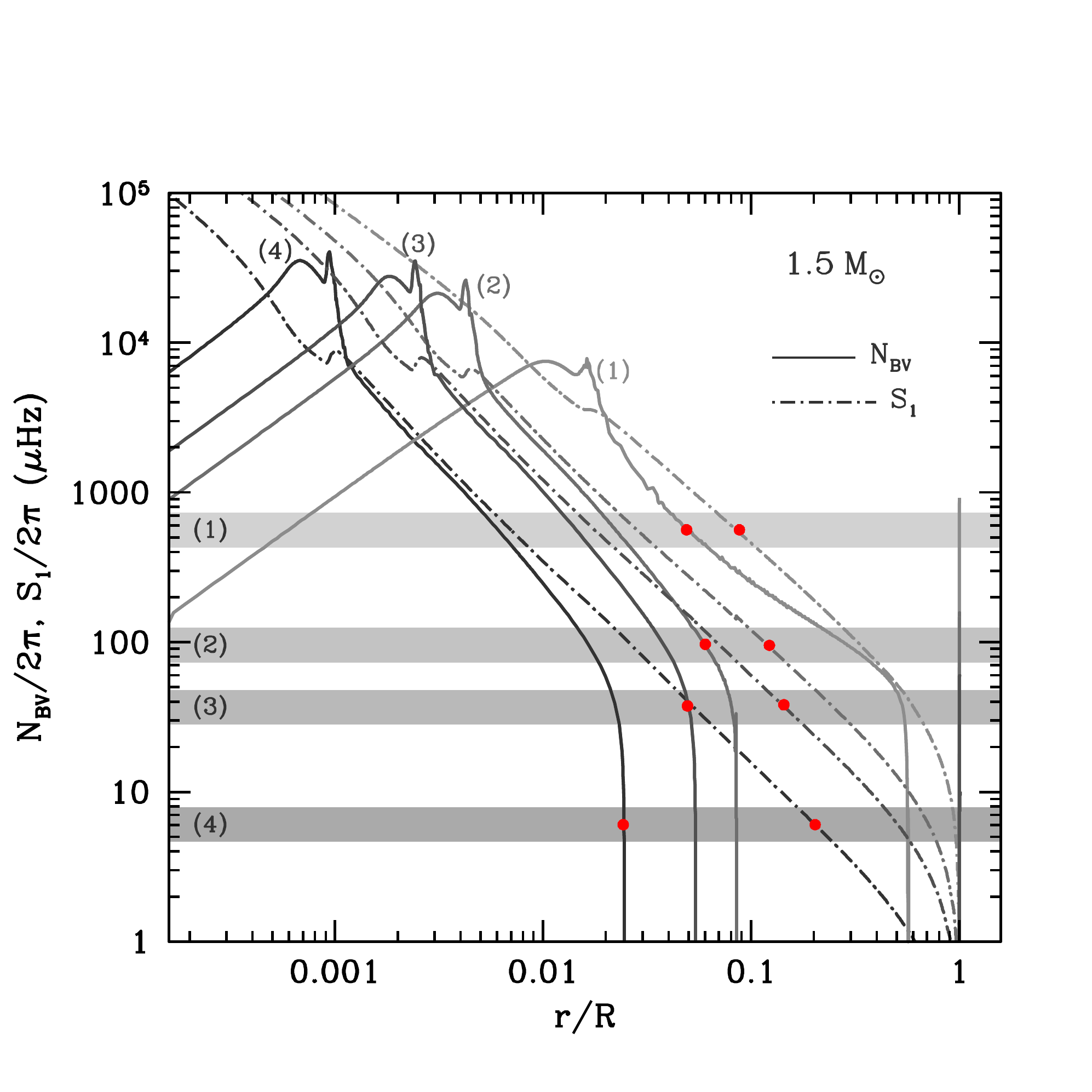}
\includegraphics[width=0.7\textwidth]{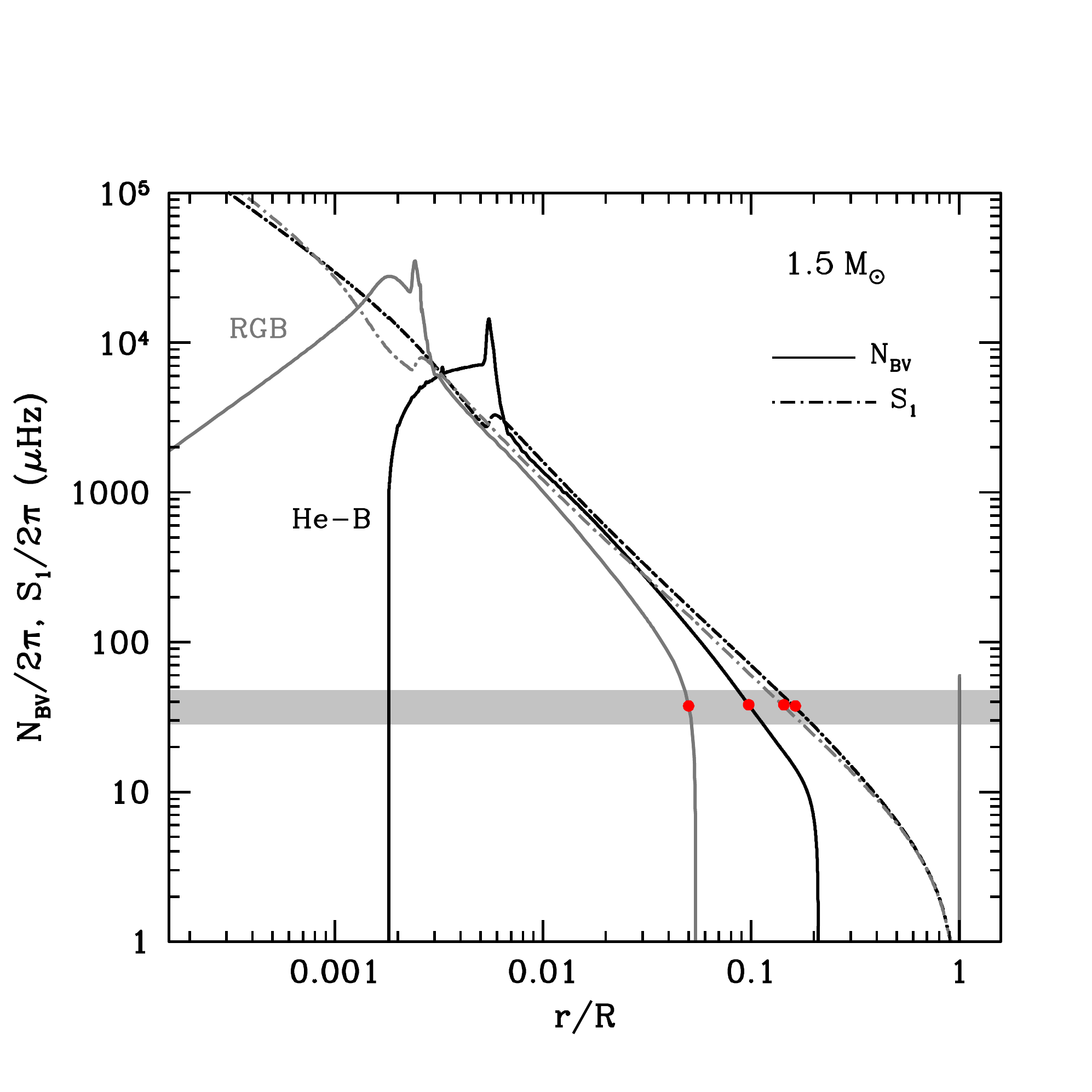}
 \caption{Propagation diagram for a 1.5~$M_\odot$ model at different
   evolutionary states. Horizontal bands correspond to the frequency
   domain of solar oscillations for each state and red dots indicate
   the limits of the respective evanescent regions. {\sl Top:}  (1)
   model at the end of the subgiant phase (2.85~$R_\odot$); (2), (3), and (4) RGB models
    with radii of 8, 12, and 30 $R_\odot$ respectively. Note
   that the size of the evanescent region increases as the radius increases.
    {\sl  Bottom:} propagation diagrams for two models with radius $\sim
   12~R_\odot$, one in the ascending RGB, the other
   in the red clump (He-B). Note that, while these models have very
   similar Lamb frequencies, the sizes of their 
   evanescent regions are quite different.}
   \label{evolN}
   \end{center}
 \end{figure}

As the star evolves in the post-MS and ascends the RGB,  the inert He-core 
continues contracting, while the H-rich envelope
expands. All that involves an increase of the Brunt-V\"ais\"al\"a frequency in the central regions 
(hence of $g$-mode frequencies, $\nu_{\rm g}\propto \Pi_0^{-1}$, Eq.~\ref{pi0}), and the drop of the mean density, and
therefore of the frequency of acoustic modes (Eq.~\ref{asym}). 
Modes with frequencies in the solar-like domain can propagate in the
gravity and acoustic cavities, and present a mixed gravity-pressure
character.  The $N_{\rm BV}$-increase close to the centre also involves
a higher density of g modes  propagating in the $g$-cavity, 
$n_{\rm g}\propto [\ell(\ell+1)]^{1/2}\int N_{\rm BV}/r\ dr$ (see Eq.~\ref{pi_spacing}).
 So, while the solar-like spectra of subgiant pulsators  are made up
 of acoustic modes and of a small number (increasing with evolution) of
 mixed modes, those of red giants  can present, in addition to radial
 modes, a large number of non-radial $g-p$ mixed modes in-between two
 radial modes \citep{2001MNRAS.328..601D, 2004SoPh..220..137C}.

The large frequency separation of radial modes decreases as the star
expands, but the value of $\langle\Delta\nu\rangle$ solely  does not
allow us to distinguish stars having the same mass and radius, but located
 in the  ascendant RGB, descendent  RGB, or central He-burning phases. 
It is the behaviour of the $g-p$ mixed modes that informs us on the inner regions of the star and potentially on its evolutionary state.  

\begin{figure}[!hpt]
 \begin{center}
\includegraphics[width=0.7\textwidth]{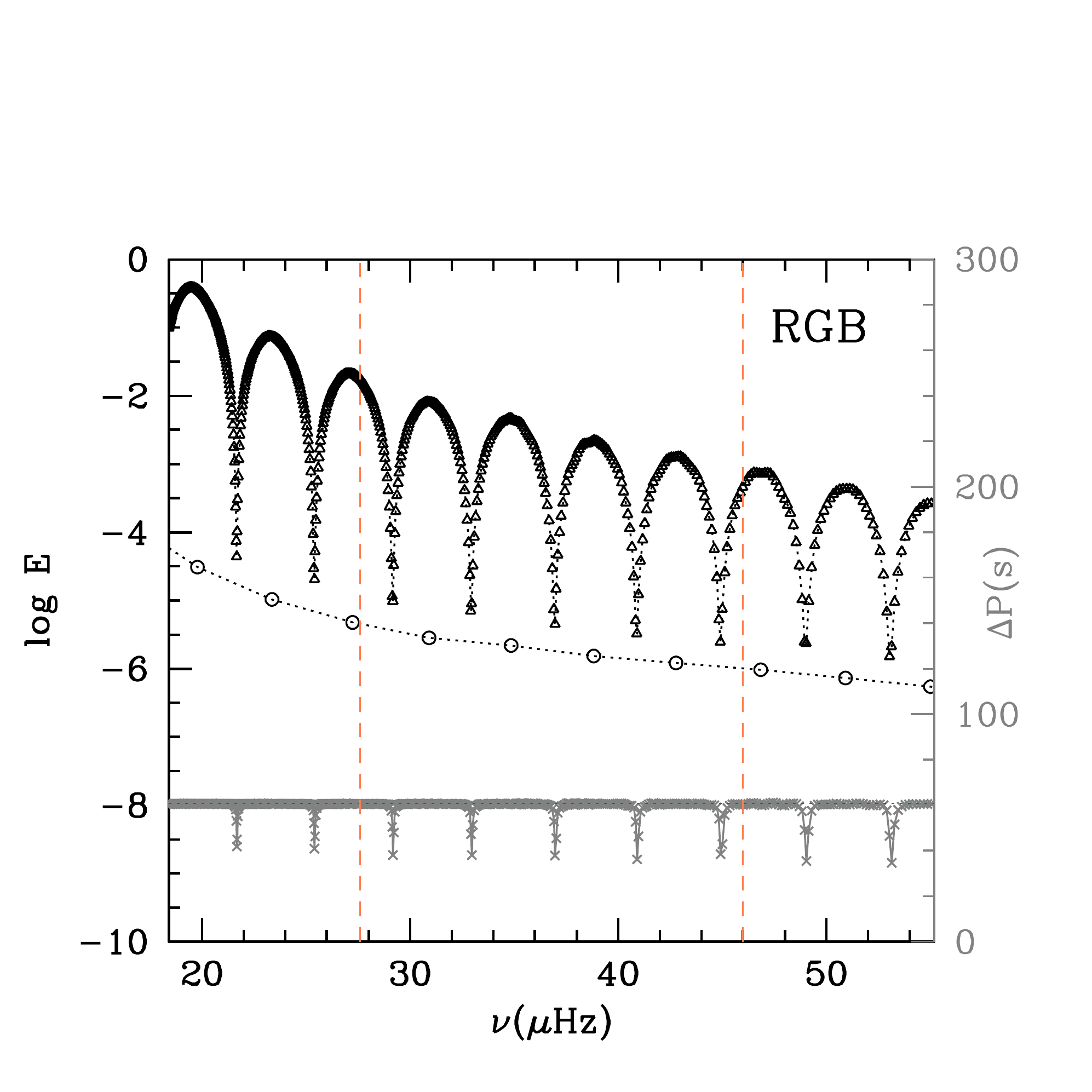}
\includegraphics[width=0.7\textwidth]{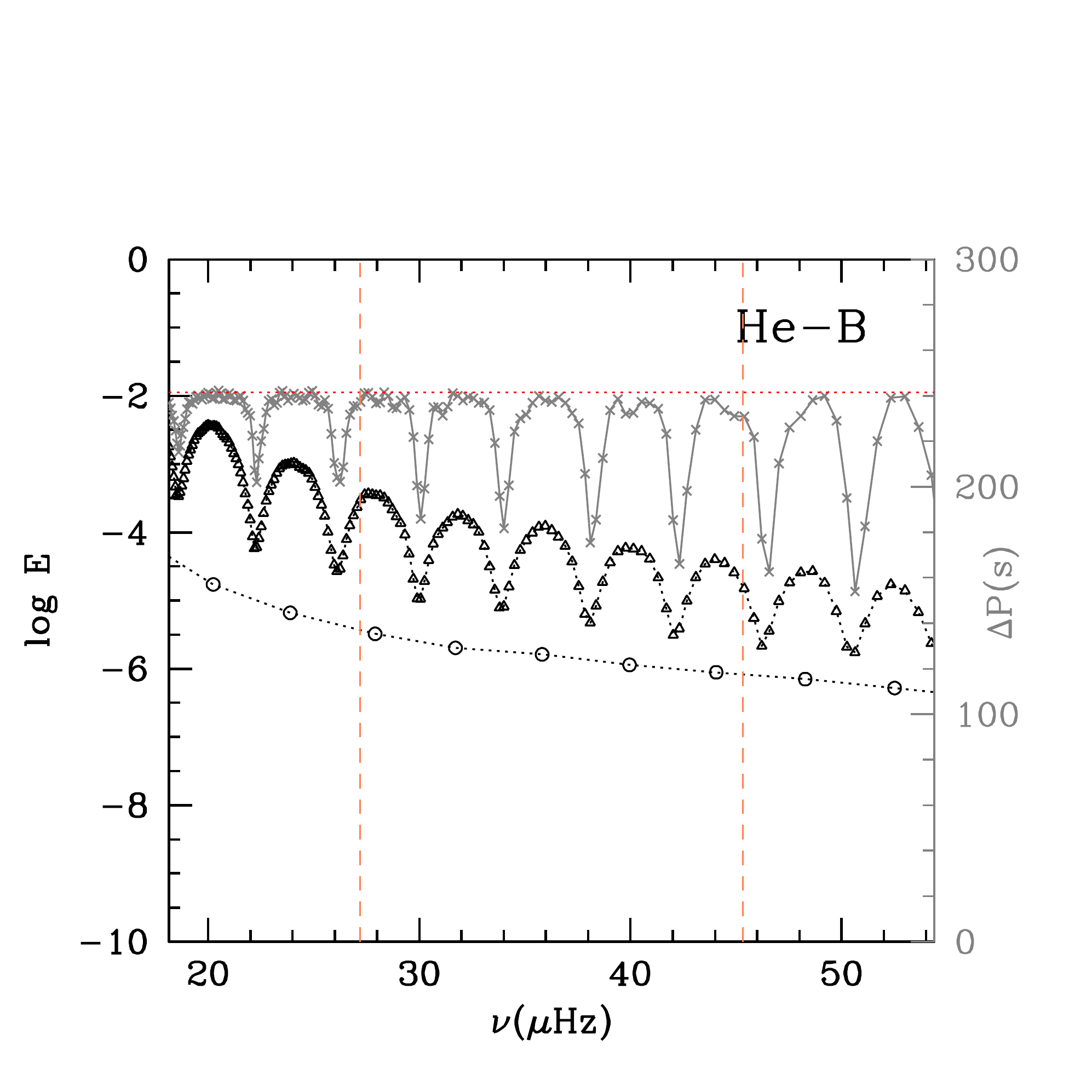}
 \caption{Oscillation spectra for models in the right panel of
   Fig.~\ref{evolN}. {\sl Top:} RGB model. {\sl Bottom:} core
   He-burning model. In each panel we plot the mode inertia ($E$)
   versus frequency for radial ($\ell=0$, circles) and dipolar
   ($\ell=1$, triangles) modes. Vertical dashed lines delimit the
   frequency domain of solar-like oscillations. Grey crosses and lines
 represent the period separation (right axis) between consecutive
 dipole modes versus frequency, and horizontal thin-lines indicate
 the value of the corresponding asymptotic period spacing. 
[Adapted
 from \citet{2013EPJWC..4303002M}.] 
 }
   \label{inertia}
   \end{center}
 \end{figure}

The dominant $p$- or $g$-character of these modes depends on the cavity
where they mainly propagate: inner region,  $g$-dominated;  envelope,
$p$-dominated or pure acoustic modes.  The dominant character may be estimated from the value of the normalized mode inertia \citep[$E$, see \eg,][and references therein]{2004SoPh..220..137C}:
\begin{equation}
E \; = \; \frac{1}{M\,|\xi|^2_{\rm ph}} \, 
\int_V \, \rho \, |\xi|^2 \, dV \quad ,
\end{equation}
\noindent
where $\rho$ is the local density. The integration is performed over the
volume $V$  of the star of total mass $M$, and ph refers to the
value of the displacement at the photosphere. Modes trapped in
high-density regions ($g$ modes) have high $E$, while pure $p$ modes such
as the radial ones have the lowest $E$. Depending on the coupling
between the two cavities, non-radial modes may be: {\it (i)} well trapped in the
acoustic cavity,  with  inertia close to
that of the radial modes and behaving as acoustic modes ($p$-dominated); {\it (ii)} well
trapped in the gravity cavity,  with very high inertia and  behaving
as pure g modes; or, {\it (iii)}  can have a significant
amplitude in both cavities, being $p-g$ mixed modes whose $g$-dominant character
increases with $E$ 
\citep{2001MNRAS.328..601D, 2004SoPh..220..137C, 2009A&A...506...57D, 2010ApJ...721L.182M,2013A&A...549A..75G}. 
As shown in
Fig.~\ref{inertia}, between two consecutive radial modes, non-radial modes are found 
 whose inertia $E$ may change by several orders of magnitude.

\begin{figure}[!ht]
 \begin{center}
\includegraphics[width=0.495\textwidth]{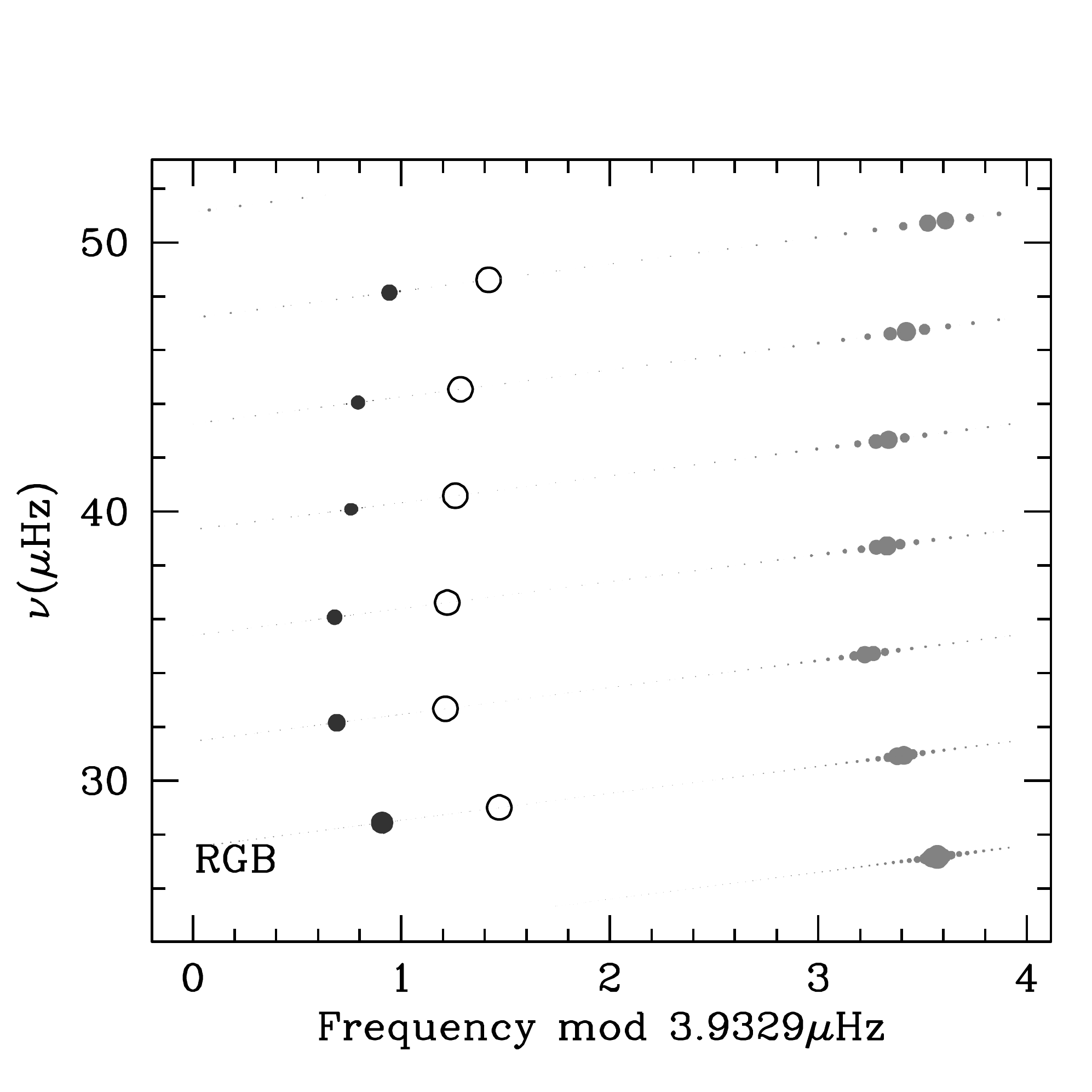}
\includegraphics[width=0.495\textwidth]{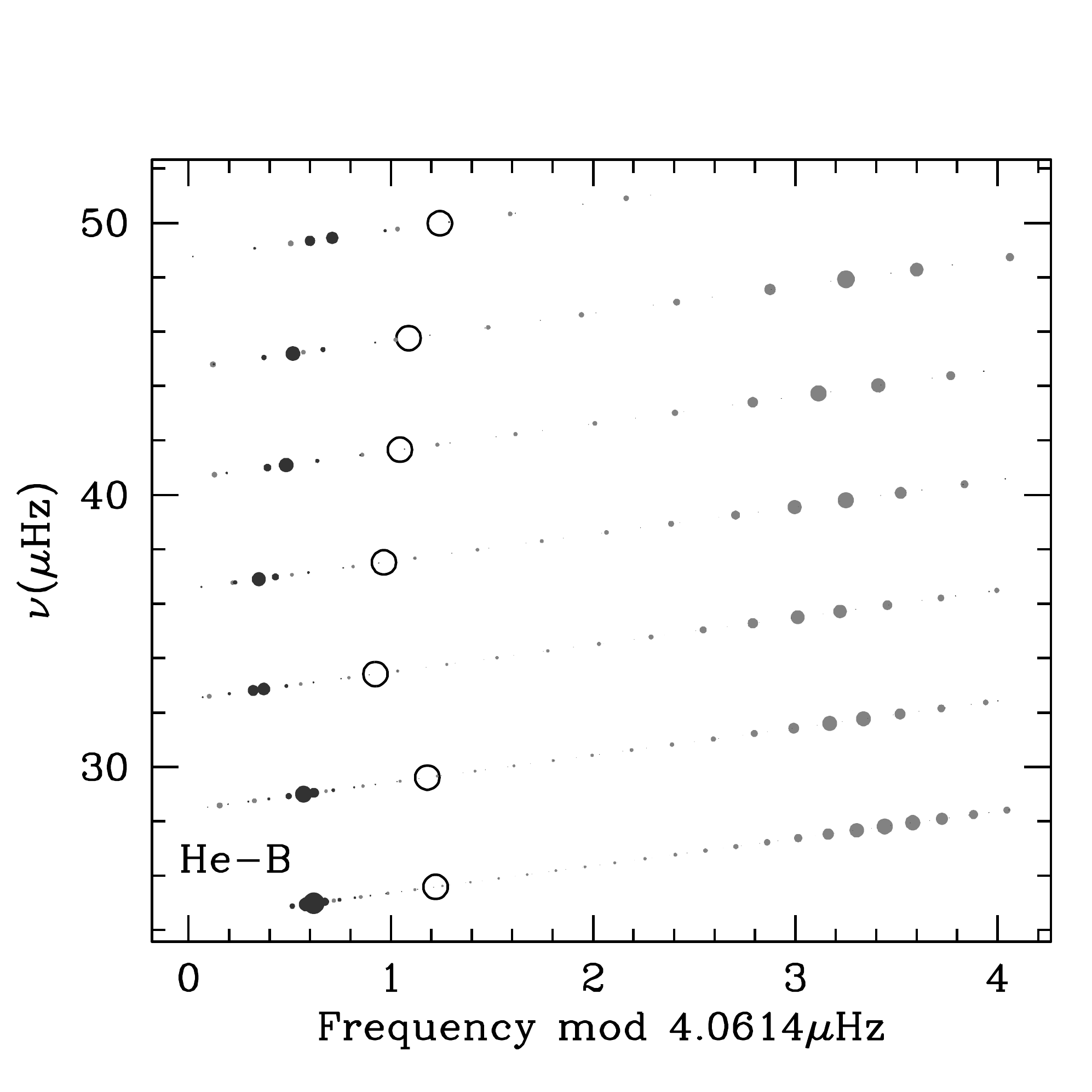}
 \caption{Expected \'echelle diagram for the oscillation spectra in
   Fig.~\ref{inertia}. Open circles correspond to $\ell=0$ modes,
   dark-solid dots to $\ell=2$ ones, and grey-solid dots to dipole
   modes. The symbol size is related to the relative amplitude
   ($\propto E^{-1/2}$) taking as reference that of the radial
   modes. {\sl Left:} RGB model. {\sl Right:} He-burning
   model. [Adapted  from \citet{2013EPJWC..4303002M}.]}
  \label{spec_rg}
   \end{center}
 \end{figure}

The amplitude of modes at different frequencies in the oscillation
spectrum results from the balance between excitation and damping rates
\citep{2001MNRAS.328..601D, 2002MNRAS.336L..65H, 2009A&A...506...57D}.
 Nevertheless an estimate of the relative amplitude of
different modes can be provided by the inverse of the square root of
the normalized inertia \citep{1999A&A...351..582H,
  2004SoPh..220..137C}.  Based on that, modes with dominant
$p$-character are expected to be more easily observed than gravity
dominated ones. Nevertheless,  as the observation time of red giants has
increased, it has become possible to  detect, around the
$p$-dominated  modes, a forest of $g$-dominated modes with a regular distribution pattern
 \citep{2011Sci...332..205B, 2011Natur.471..608B,  2011A&A...532A..86M}. For pure $g$-modes
this regularity would correspond to the asymptotic period spacing
($\langle \Delta P\rangle_\mathrm{asym}$) (see Sect.~\ref{period-spacing}).
The value of the measured period spacing ($\langle\Delta P\rangle$) is  different from the asymptotic
one since the modes involved in the estimation  are actually    $g$-mixed modes
whose behaviour deviates from that of pure $g$-modes.  Nevertheless,
\citet{2012A&A...540A.143M} showed that  it is possible to recover
$\langle \Delta P\rangle_\mathrm{asym}$ from  $\langle\Delta P\rangle$ and
hence acces to stellar core properties.

\begin{figure}[!ht]
\begin{center}
\includegraphics[width=0.99\textwidth]{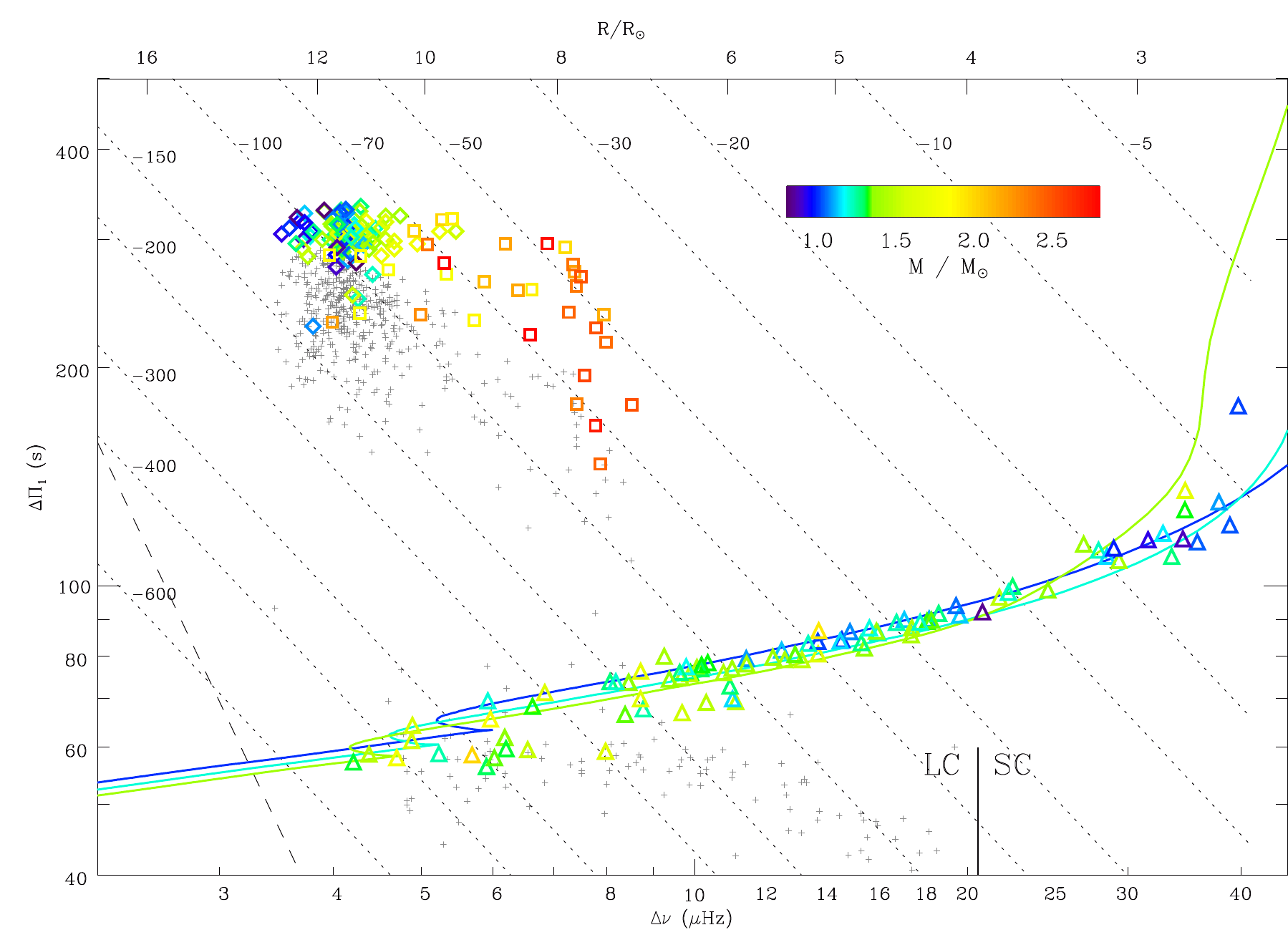}
\caption{Gravity-mode period spacing 
 as a function of the pressure-mode large frequency spacing. 
 RGB stars are indicated by triangles; clump stars by diamonds; 
 secondary clump (He-burning stars  in the transition
 between low- and intermediate- mass stars)  by squares.
   The seismic estimate of the mass is given by the colour code. 
     The solid coloured lines correspond to a grid of stellar models with masses of
      $1, 1.2$, and $1.4 M_\odot$,
from the ZAMS to the tip of the RGB. [From  \citet{2012A&A...540A.143M}.] 
}
\label{dpdv_mosser}
\end{center}
\end{figure}

The coupling between the two cavities depends on the density contrast
between the core and the envelope, and hence on the evolutionary state
and stellar mass.

As the star climbs the RGB and the inert He-core contracts, the peak
of $N_{\rm BV}$ shifts more and more inwards (see Fig.~\ref{evolN}), and the 
evanescent region, or the potential barrier that separates the two
cavities ($\int K_{r} \, dr$, Eq.~\ref{muelle}), becomes wider and makes
the trapping of modes in each cavity more efficient: as the star goes
from the bottom to the tip of the RGB, the amplitude of the numerous
$g$-dominated modes decreases, as well as the period separation between
consecutive g dominated modes ($n_{\rm g}$ increases).  Once the
conditions for He-ignition are reached, the onset of nuclear
burning is accompanied by the expansion of the central regions and the
development of a convective core. Both effects act in the same sense:
a lower $N_{\rm BV}$ maximum value located at larger radius and the
small convective core, involve a decrease of $n_{\rm g}$ and an
increase of $\langle \Delta P\rangle_\mathrm{asym}$. Moreover, the lower density
contrast ($\rho_{\rm c}/\langle \rho \rangle$) between core and
envelope implies a stronger coupling between the
two cavities. Figures \ref{evolN},  \ref{inertia}, and
\ref{spec_rg}  compare the structure and
seismic properties of two models of $1.5\,M_{\odot}$ with the same
radius and different evolutionary state (RGB  and  red clump --He-B):  they have
the same $\nu_{\rm max}$ and $\langle
\Delta\nu\rangle$, but $\rho_{\rm c}/\langle \rho \rangle$  is 10
times smaller in the He-B model than in the RGB one. That implies a
larger ($\sim 4$ times) period spacing between consecutive $g$-modes and
a higher $p-g$ mixed character of modes in the He-B model. While the RGB spectrum shows a
behaviour close to that of MS solar-like pulsators where only
acoustic modes appear,  the He-B one shows a very scattered 
$\ell=1$ ridge.

The scatter of dipole-modes ridge (together with the values of $\Delta\nu$ and
 $\nu_{\rm max}$) appears then  as a tool to
 distinguish between RGB and clump stars \citep{2010ApJ...721L.182M}.
 Moreover, at a given $\langle\Delta\nu\rangle$,  the difference of
 $\Delta P $-values splits the $G-K$ red-giant pulsators in two groups: stars with $\Delta P > 100$~s
 (He-burning stars) and stars with $\Delta P < 60~$s (RGB)
 \citep{2011Natur.471..608B,  2011A&A...532A..86M, 2012A&A...540A.143M}. The evolutionary state of red giants can  then be 
 characterised by two global properties of the oscillation spectra:
 $\langle\Delta\nu\rangle$ and $\langle \Delta P \rangle$, see Fig.~\ref{dpdv_mosser}.

Coming back to Fig.~\ref{mass-age}, the possibility to discriminate between the
evolutionary states of red giants implies that, once the metallicity
has been derived, the uncertainty on their ages becomes  of the order
15\%, to be compared with the high uncertainties affecting classical
methods such as isochrone fitting.

 Actually the detection of $g$-dominated modes in RGs has revealed
 other  important aspects of their evolution and structure that may  directly or indirectly affect age-dating. In
 particular it has been possible to estimate their internal rotation
 profiles and it turned out that those predicted by current models 
are at odds with those deduced from  seismic observations
\citep{2012A&A...544L...4E,  2012ApJ...756...19D, 2012A&A...548A..10M, 2013A&A...549A..74M, 2014A&A...564A..27D},
  This indicates that transport of angular momentum and therefore 
more importantly here the associated chemical mixing
are not properly modelled. This must have 
 some impact on the age-dating, which remains to be quantified. 
As a prospect,  the ability of probing the core of
 RGs with  $g$-dominated modes will certainly help  to get insights in the transport
 mechanisms to take into account both during the  red giant  phase 
 and  prior to that stage,  on the MS \citep{2013ApJ...766..118M}. 
 This in turn will improve the accuracy of age determination
 based on stellar modelling. 
 
\section{Conclusion}

When oscillation frequencies or derived quantities like frequency separations or separation ratios are taken as constraints for stellar 
modelling, the uncertainties on the determination of the age, mass, and radius of stars decrease spectacularly with respect to 
what is obtained from the constraints on classical parameters. 

For main-sequence solar-like oscillators, considering different possible sets of model input physics and model free parameters, the 
uncertainty on age may reach $90$ per cent when solely the classical parameters constrain the models. When seismic constraints are 
added, the uncertainty on age drops to a level of about $10$ per cent as in the case of HD 52265. For subgiants stars, and for a fixed 
set of input physics, the uncertainty typically drops from $35-40$ per cent to $5-7$ per cent. Finally, seismology has 
given the first very reliable clues about  the mass and evolutionary stage of red giant stars, which cannot be distinguished on the basis 
of solely the classical stellar parameters.

Further progress still requires to better understand the physics governing stellar structure and evolution, as 
well as to improve the frequency measurements to fully exploit asteroseismic diagnostics, and last but not least, to go on improving 
the accuracy on classical parameters. In that respect,  a very high precision on the determination of the stellar classical parameters
will be provided by the \textsl{Gaia}-ESA mission \citep{2001a&a...369..339p}, 
but to  achieve a satisfactory determination of stellar ages, 
seismic data  such as expected from the \textsl{PLATO}-ESA mission \citep{2013arxiv1310.0696r} 
will be essential.

\section*{Acknowledgements}
The authors would like to thank the ``Formation permanente du CNRS'' for 
financial support.  
The preparation and writing of these lectures largely benefited from the
 use of the SIMBAD database,
operated at CDS, Strasbourg, France and of the NASA's Astrophysics Data 
System.


\end{document}